\documentclass[lettersize,journal]{IEEEtran}
\usepackage{amsmath,amsfonts}
\usepackage{algorithmic}
\usepackage{algorithm}
\usepackage{array}
\usepackage[caption=false,font=footnotesize,labelfont=rm,textfont=rm]{subfig}
\usepackage{textcomp}
\usepackage{stfloats}
\usepackage{url}
\usepackage{verbatim}
\usepackage{graphicx}
\usepackage{cite}
\usepackage{multirow}
\usepackage{amssymb}
\usepackage{xcolor}
\begin{document}

\title{Quality-of-Service Aware LLM Routing for Edge Computing with Multiple Experts}

\author{Jin Yang, Qiong Wu, Zhiying Feng, Zhi Zhou,~\IEEEmembership{Member,~IEEE}, \\
Deke Guo,~\IEEEmembership{Senior Member,~IEEE}, and Xu Chen,~\IEEEmembership{Senior Member,~IEEE}
\thanks{Corresponding Authors: Xu Chen and Qiong Wu.}
\thanks{Jin Yang, Zhiying Feng, Zhi Zhou, and Xu Chen are with the School of Computer Science and Engineering, Sun Yat-sen University, Guangzhou, Guangdong 510006, China (e-mail: yangj557@mail2.sysu.edu.cn; fengzhy26@mail2.sysu.edu.cn; zhouzhi9@mail.sysu.edu.cn; chenxu35@mail.sysu.edu.cn).}
\thanks{Qiong Wu is with the Department of Computer Science and Engineering, The Hong Kong University of Science and Technology, Hong Kong, China (e-mail: cseqiongwu@ust.hk).}
\thanks{Deke Guo is with the School of Computer Science and Engineering, Sun Yat-sen University, Guangzhou, Guangdong 510006, China, and was with the College of Systems Engineering, National University of Defense Technology, Changsha, Hunan 410073, China (e-mail: guodk@mail.sysu.edu.cn).}

}



\maketitle

\begin{abstract}
Large Language Models (LLMs) have demonstrated remarkable capabilities, leading to a significant increase in user demand for LLM services. However, cloud-based LLM services often suffer from high latency, unstable responsiveness, and privacy concerns. Therefore, multiple LLMs are usually deployed at the network edge to boost real-time responsiveness and protect data privacy, particularly for many emerging smart mobile and IoT applications. Given the varying response quality and latency of LLM services, a critical issue is how to route user requests from mobile and IoT devices to an appropriate LLM service (i.e., edge LLM expert) to ensure acceptable quality-of-service (QoS). Existing routing algorithms fail to simultaneously address the heterogeneity of LLM services, the interference among requests, and the dynamic workloads necessary for maintaining long-term stable QoS. To meet these challenges, in this paper we propose a novel deep reinforcement learning (DRL)-based QoS-aware LLM routing framework for sustained high-quality LLM services. Due to the dynamic nature of the global state, we propose a dynamic state abstraction technique to compactly represent global state features with a heterogeneous graph attention network (HAN). Additionally, we introduce an action impact estimator and a tailored reward function to guide the DRL agent in maximizing QoS and preventing latency violations. Extensive experiments on both Poisson and real-world workloads demonstrate that our proposed algorithm significantly improves average QoS and computing resource efficiency compared to existing baselines.
\end{abstract}


\section{Introduction}
\label{sec:intro}

\begin{figure}[tbp!]
\centerline{\includegraphics[width=0.45\textwidth]{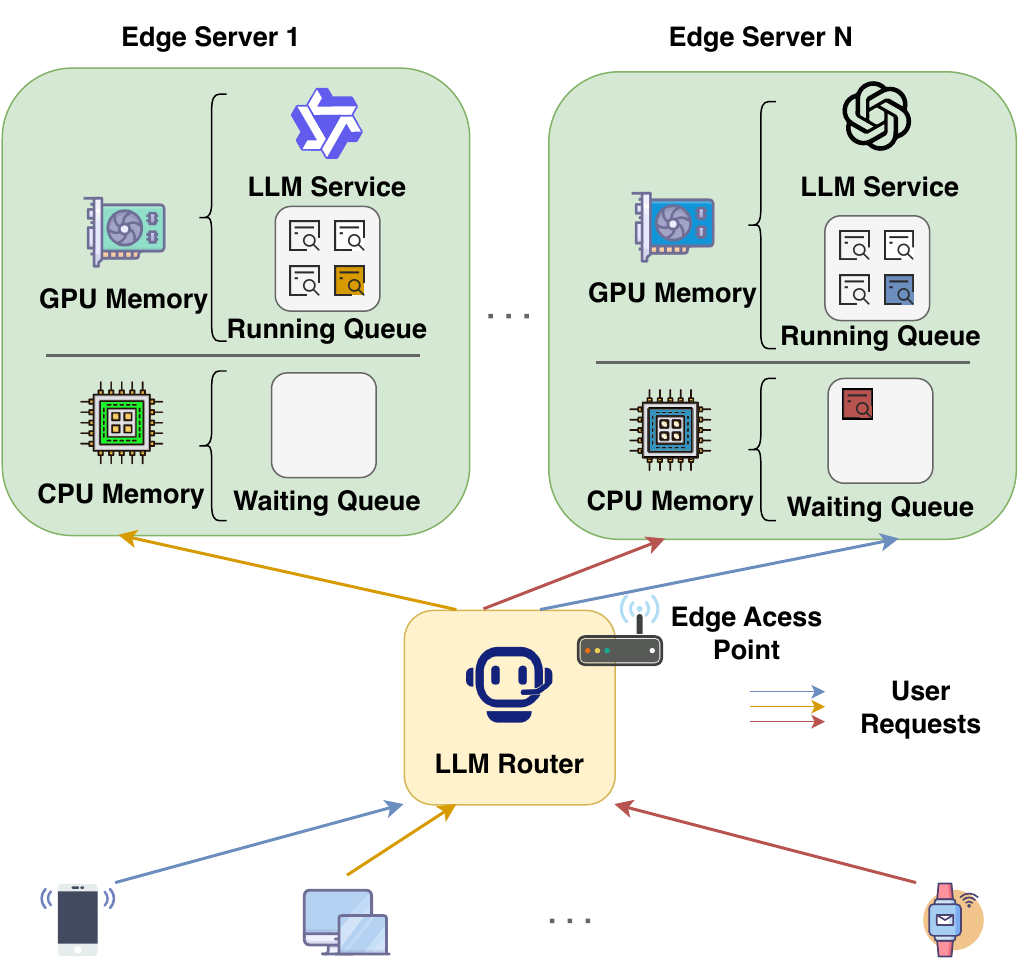}}
\caption{LLM routing at the edge with multiple experts.}
\label{fig:infrastructure}
\end{figure}

\IEEEPARstart{R}{ecently}, innovative smart mobile and IoT applications based on large language models (LLMs), such as smart-home AI assistants ~\cite{10729865} and intelligent robots ~\cite{ijcai2024p844}, have garnered significant attention worldwide and greatly enhanced convenience and efficiency in people's daily lives and work. Currently, these applications mainly rely on the LLM services deployed on the cloud to serve user inference requests originating from resource-limited devices (e.g., smartphones, laptops, and wearable devices) at the network edge. However, this common practice faces several critical issues: (i) significant latency due to the long transmission distance between the cloud and edge devices ~\cite{yang2023edgefm}, (ii) unreliable performance due to fluctuating network conditions, especially over wireless networks ~\cite{wdmoe}, (iii) privacy concerns arising from the transmission of sensitive user data over public networks ~\cite{yan2024protecting}.

Edge computing, a distributed computing paradigm that extends cloud capabilities to the network edge, is a promising solution to mitigate these issues by reducing response times and safeguarding data privacy ~\cite{zhou2019edge}. In this context, deploying multiple heterogeneous LLM services at the edge emerges as a highly effective strategy ~\cite{du2023enabling,wang2024toward}. As depicted in Figure \ref{fig:infrastructure}, LLM services are positioned at the edge, acting as edge experts. To flexibly accommodate the diverse needs of a wide range of users, these edge experts typically possess distinct expertise ~\cite{llm_blender}. Moreover, these edge experts may have different CPU and GPU memory resources, leading to significant heterogeneity. User requests from resource-constrained edge devices are first directed to an Edge Access Point (eAP), which is connected to these edge experts via a Local Area Network (LAN). Since the heterogeneity of edge experts, they provide various response quality and latency. To ensure acceptable QoS, the LLM router in the eAP routes these requests to the most appropriate edge expert. However, achieving a satisfactory QoS for user requests across multiple edge experts for practical applications presents the following challenges.

\begin{figure}[t!]
\centerline{\includegraphics[width=0.45\textwidth]{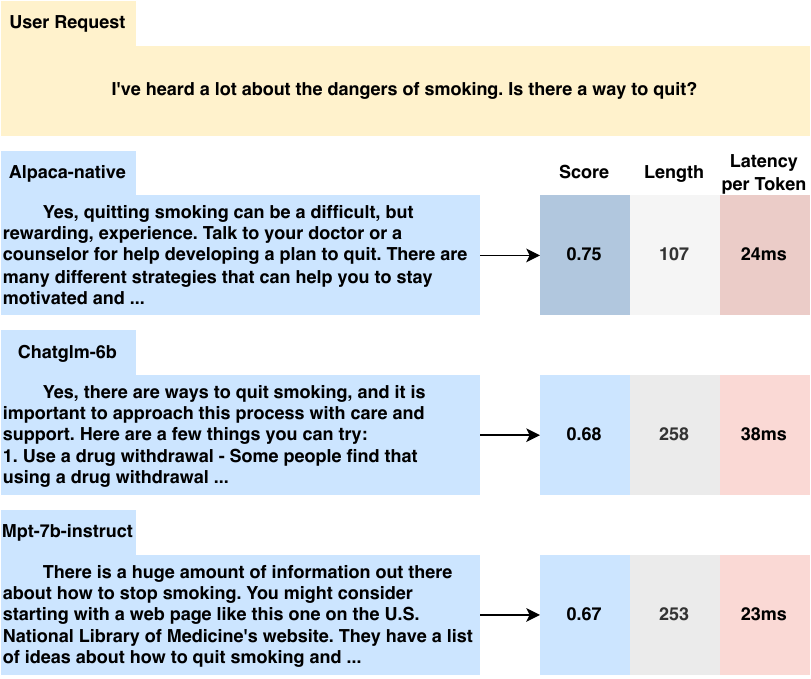}}
\caption{An example of the LLM services. Given a user request, different LLM services exhibit varying response quality, response length, and latency per token.}
\label{fig:pre1}
\end{figure}

\textbf{Heterogeneity of LLM services.} \textcolor{black}{Due to differences in training data and model architectures, LLM services exhibit substantial heterogeneity in their service capabilities \cite{llm_blender}. As illustrated in Figure \ref{fig:pre1}, different LLM services demonstrate markedly different response quality, response length, and latency per token when handling the same user request. This heterogeneity presents a fundamental challenge for existing LLM request scheduling systems \cite{jain2024intelligent}, which typically assume homogeneous service capabilities. In real-world deployment scenarios, this heterogeneity can result in inefficient resource allocation and suboptimal response quality. For instance, routing requests to LLM services with longer response lengths or inferior response quality may exacerbate GPU memory pressure and degrade overall QoS. Consequently, the challenge lies in effectively harnessing this heterogeneity to optimize request routing across diverse LLM services.
} 

\textbf{Interference among requests.} \textcolor{black}{Practical LLM services are usually deployed on sophisticated inference systems like Orca \cite{orca} and vLLM \cite{vllm}, which employ advanced techniques like iteration-level scheduling. These systems are designed to serve multiple requests concurrently, thereby reducing queuing delays. As illustrated in Figure \ref{fig:infrastructure}, each edge expert maintains a running queue and concurrently processes all user requests in the running queue. However, such concurrent processing introduces interference among requests \cite{jain2024intelligent}. Existing LLM routing systems ~\cite{stripelis2024polyrouter,zooter,ding2024hybrid,routellm} fail to capture this interference effect, which significantly increases the response latency experienced by user requests and degrades the overall QoS. Therefore, effectively capturing the interference among requests is crucial for optimizing the overall QoS.}

\textbf{Dynamic workloads.} \textcolor{black}{As revealed by BurstGPT \cite{burstgpt}, real-world LLM workloads are highly dynamic, influenced by diverse behaviors of users, systems, and LLM models. As depicted in Figure \ref{fig:infrastructure}, the number of requests managed by each edge expert can vary over time, influenced by both the available computational resources and the temporal pattern of incoming requests, resulting in fluctuating workload conditions. While existing LLM routing systems ~\cite{stripelis2024polyrouter,zooter,ding2024hybrid,routellm} effectively exploit LLM heterogeneity to optimize QoS for individual requests, they often overlook the real-time LLM workload conditions. This oversight poses a dual challenge: (i) routing new requests to already overloaded edge experts increases response latency, compromising the immediate QoS, while (ii) underutilizing available computational resources reduces long-term processing efficiency, negatively impacting the long-term QoS. Consequently, achieving balanced workload distribution across edge experts during request routing is essential for optimizing the overall QoS under dynamic LLM workloads.}

\textcolor{black}{To address these challenges, we propose a novel DRL-based QoS-aware LLM routing algorithm. Compared to existing solutions, our algorithm better maximizes the long-term QoS for user requests across heterogeneous edge experts for practical applications with dynamic LLM workloads. The technical contributions of this paper are listed as follows:}
\begin{itemize}
\item[1.] To maximize the long-term QoS across heterogeneous edge experts, we propose a novel DRL-based QoS-aware LLM routing algorithm to achieve optimized routing under dynamic LLM workloads.
\item[2.] Due to the dynamic nature of the global state, we propose a dynamic state abstraction technique to encode the dynamic global state features with a HAN, which maps the raw states into a more compact space. 
\item[3.] To guide the DRL agent in maximizing overall QoS and preventing latency requirement violations, we propose an action impact estimator and design a reward function for our DRL agent accordingly. 
\item[4.] We conduct extensive experiments on emulated Poisson workloads and real-world LLM serving workloads to validate the effectiveness of our algorithm. Experimental results show its superiority in long-term QoS and resource efficiency over baselines.
\end{itemize}

The remainder of this paper is organized as follows. Section \ref{sec:related} summarizes related works. Section \ref{sec:prelim} introduces preliminary background on LLM inference and illustrates the heterogeneity of LLM services and the interference among requests. Section \ref{sec:problem} describes our scenario and formulate our QoS-aware LLM routing problem. Section \ref{sec:algo} first proposes the DRL-based QoS-aware LLM router, then details the dynamic state abstraction technique and the QoS-aware reward design. Section \ref{sec:comp_analy} presents the computational complexity analysis of the proposed routing algorithm. We conduct extensive experiments to show the superiority of our algorithm in Section \ref{sec:evaluation}. Section \ref{sec:conclusion} gives the conclusion.

\section{Related Work}
\label{sec:related}

\textbf{LLM Serving Algorithms.} Numerous LLM serving systems are proposed to address the unique challenges of LLMs. Orca \cite{orca} introduced an iteration-level scheduling strategy to schedule batch execution of user requests at the iteration level, significantly improving the throughput of the inference system. vLLM \cite{vllm} introduced the PagedAttention algorithm, achieving more efficient management of key-value caches and considerably reducing the memory footprint during LLM inference. To address the head-of-line blocking issues and improve interactive LLM serving efficiency, Qiu et al. \cite{qiu2024efficient} proposed a speculative shortest-job-first (SSJF) scheduler that leverages a lightweight proxy model to predict LLM output sequence lengths. Similarly, $S^3$ \cite{jin2023s} system predicts output sequence lengths and schedules generation requests accordingly, increasing resource utilization and performance. FlexGen \cite{sheng2023flexgen} introduced a high-throughput generation engine for running LLMs with limited GPU memory, which can be flexibly configured under various hardware resource constraints by aggregating memory and computation from the GPU, CPU, and disk. Some other systems focus on GPU kernel optimization and kernel fusion \cite{dao2022flashattention}, model parallelism \cite{xuanlei2024hetegen, oh2024exegpt}, batching algorithm \cite{orca, qiu2024efficient, jin2023s}, KV-cache management \cite{vllm, gao2024attentionstore} and disaggregated inference \cite{patel2024splitwise, zhong2024distserve}. However, these systems focus on optimizing aggregated server-side performance, often failing to consider the long-term QoS provided for users.

\textbf{LLM Routing Algorithms.} Recent works introduce LLM routing algorithms, aiming to select the best LLM for specific user inputs before inference \cite{lu2024merge}. Shnitzer et al. \cite{shnitzer2023large} take the lead in exploring the feasibility and limitations of learning routers using various benchmark datasets. Octopus-v4 \cite{octopusv4} introduced a router model leveraging functional tokens to intelligently direct user requests to the most appropriate vertical model and reformat the query to achieve the best performance. Zooter \cite{zooter} introduced a reward-guided routing method distilling rewards on training requests to train a routing function, which can distribute each query to the LLM with expertise about it. To generalize across new LLMs and different tasks, GraphRouter \cite{feng2024graphrouter} introduced a novel inductive graph framework that fully utilizes the contextual information among tasks, queries, and LLMs to enhance the LLM routing process. However, these works focus primarily on achieving the best response quality while neglecting response latency. 

Some recent works have started to address response latency alongside response quality. Hybrid LLM \cite{ding2024hybrid} employed a router that dynamically assigns queries to either a small or large model based on the predicted query difficulty and a tunable desired quality level, allowing for a flexible trade-off between quality and cost according to the specific scenario requirements. RouteLLM \cite{routellm} trains routers using human preference data and data augmentation techniques to enhance performance, optimizing the balance between cost and response quality by dynamically selecting between a stronger and a weaker LLM during inference. Eagle \cite{zhao2024eagle}, a novel LLM routing approach that combines global and local ELO ranking modules, overcomes scalability and real-time adaptation challenges by evaluating both general and specialized LLM abilities, providing a scalable, training-free solution that enhances LLM selection quality and reduces computational overhead. PolyRouter \cite{stripelis2024polyrouter}, a non-monolithic LLM querying system, seamlessly integrates various LLM experts into a single query interface and dynamically routes incoming queries to the most high-performant expert based on the query's requirements, balancing cost and quality effectively. To effectively evaluate the router capability and limitations, RouterBench \cite{hu2024routerbench} posed a new benchmark mainly focusing on response quality and economic cost. However, these works do not account for the dynamic workloads and fail to optimize the long-term QoS.

\textbf{DRL Methods for Request Routing.} DRL has demonstrated its effectiveness in online decision-making, including request scheduling \cite{shyalika2020reinforcement}. Several works leverage the DRL algorithm for routing requests, such as web service requests \cite{shen2023collaborative}, machine learning tasks \cite{crankshaw2017clipper, li2023tapfinger} (e.g., image classification and speech recognition). KaiS \cite{shen2023collaborative} introduced a reinforcement learning scheduling framework for edge-cloud networks to improve the long-term throughput rate of web service request processing. Clipper \cite{crankshaw2017clipper}, a general-purpose low-latency predictive model serving system, introduced an adaptive model selection technique based on the Exp3 \cite{auer2002nonstochastic} algorithm to reduce prediction latency and enhance prediction throughput, accuracy, and robustness. TapFinger \cite{li2023tapfinger} introduced a multi-agent reinforcement learning (MARL) framework that minimizes the total completion time of machine learning tasks in a multi-cluster edge network through co-optimizing task placement and fine-grained multi-resource allocation. 
Although DRL algorithms for request scheduling have been extensively studied, strategies specifically tailored to LLM service workloads remain under-explored. Recently, Jain et al. \cite{jain2024intelligent} proposed a heuristic-guided DRL-based intelligent router for LLM workload scheduling, considering the distinct characteristics of the two phases in LLM workload. However, this work is tailored to serve homogeneous LLM instances and does not consider optimizing QoS for user requests across multiple LLM services. Additionally, this work does not incorporate fine-grained request-level features in the design of state features, which results in the loss of detailed information on each request.

\section{Preliminary}
\label{sec:prelim}

\subsection{LLM Inference}

\begin{figure}[t!]
    \centering
    \includegraphics[width=0.53\textwidth]{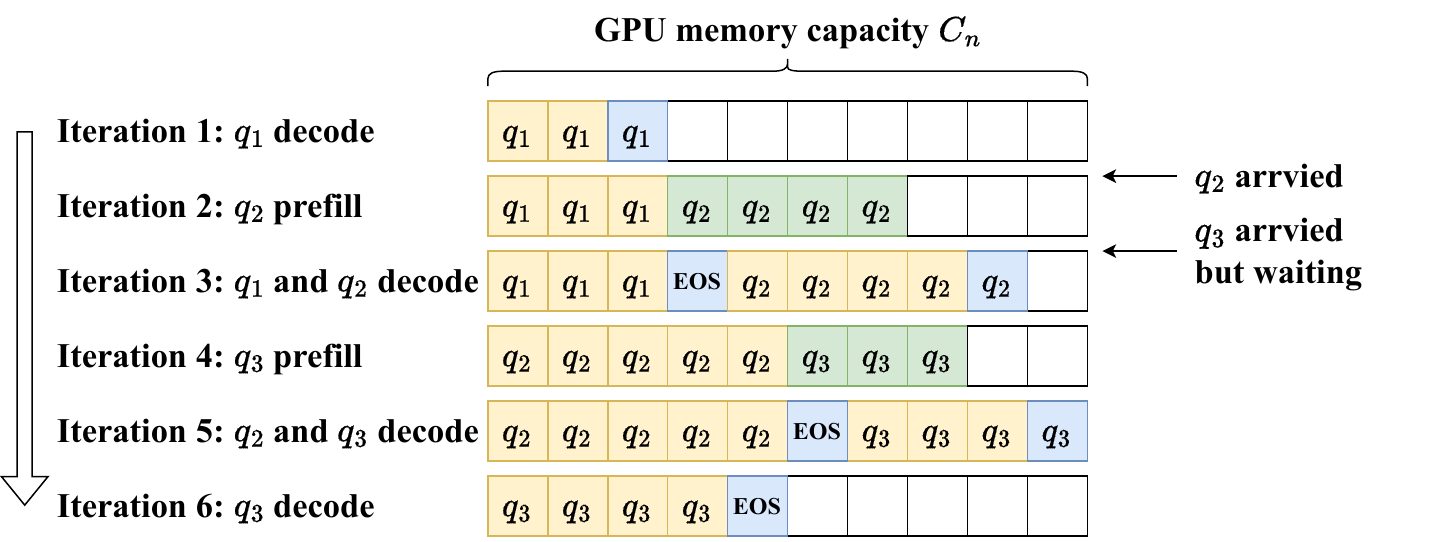} 
    \caption{An example of iteration-level scheduling for LLM inference. }
    \label{fig:inference}
\end{figure}

Generative LLM inference consists of two phases: the prefill phase and the decode phase. In the prefill phase, the model receives the prompt, a sequence of tokens $X = [x_1, ..., x_s]$ of length s, where $x_i$ denotes a token and $s$ denotes the length of the prompt. The model then computes and saves the key-value caches of each token and produces the first token $y_1$. Following this is the decode phase, where the model appends the previously generated token $y_{<i}$ to the input and auto-regressively decodes subsequent tokens. Specifically,
\begin{equation*}
    P(Y|X) = \prod_{i=1}^t g_\xi(y_{i} | y_{<i}, X),
\end{equation*}
where $Y = [y_1, ..., y_t]$ is the output response with length $t$ and $g_\xi$ refer to the LLM. The decoding step is repeated until the stop criteria are met, such as reaching the maximum token limit (e.g., we set the maximum token limit as 300) or encountering an end-of-sequence token. The computation of the decode phase is significantly reduced due to the key-value caches. Specifically, all the previous tokens do not need to pass any linear layers in the model. Due to the auto-regressive decoding process in LLM inference, the generative pattern $P(Y|X)$ and the output response $Y$ can vary based on the prompt $X$ and the LLM $g_\xi$. This variability can lead to differences in response quality and response length.

To mitigate queuing delays, iteration-level scheduling techniques, as introduced in \cite{orca}, are often employed to process requests concurrently. As depicted in Figure \ref{fig:inference}, the edge expert manages the running queue in each iteration, optimizing GPU memory utilization and computing power. Upon the arrival of a new request, if sufficient GPU memory is available, the edge expert performs a prefill operation for this request in the current iteration—saving its key-value cache and seamlessly integrating it into the running queue (e.g., in iteration 2, the newly arrived request $q_2$ processes the prefill phase and is added to the running queue). However, should GPU memory be insufficient, the incoming request will wait until space becomes available following the completion of other requests. Only then can the prefill operation be executed, and the request is subsequently added to the running queue (e.g., $q_3$, which had to await an additional iteration until $q_1$ finished and freed up memory before it could integrate into the running queue). In iterations where no new requests need to integrate into the running queue, the edge expert decodes the existing requests in parallel, efficiently utilizing computational resources.

\subsection{LLM Services Heterogeneity}
\label{sec:heter}

\begin{figure}[t!]
    \centering
    \subfloat[Response scores]{
		\includegraphics[width=0.23\textwidth]{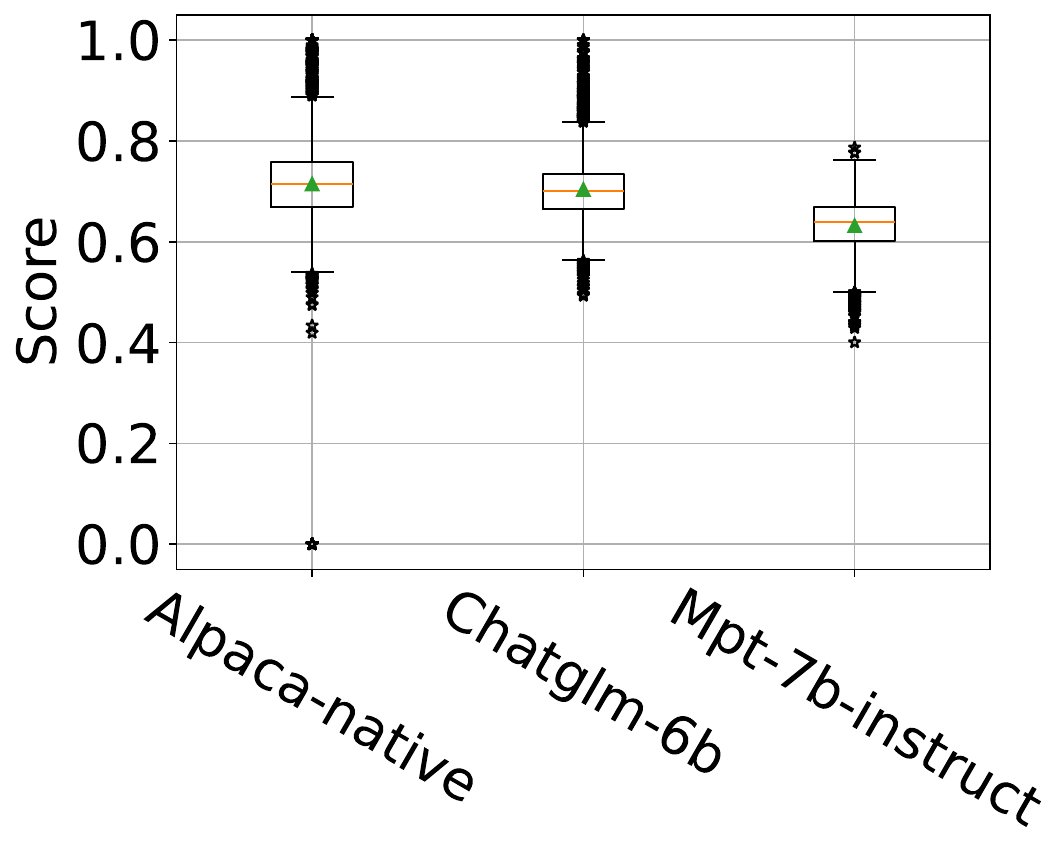}
    }
    \hfil
    \subfloat[Response lengths]{
		\includegraphics[width=0.23\textwidth]{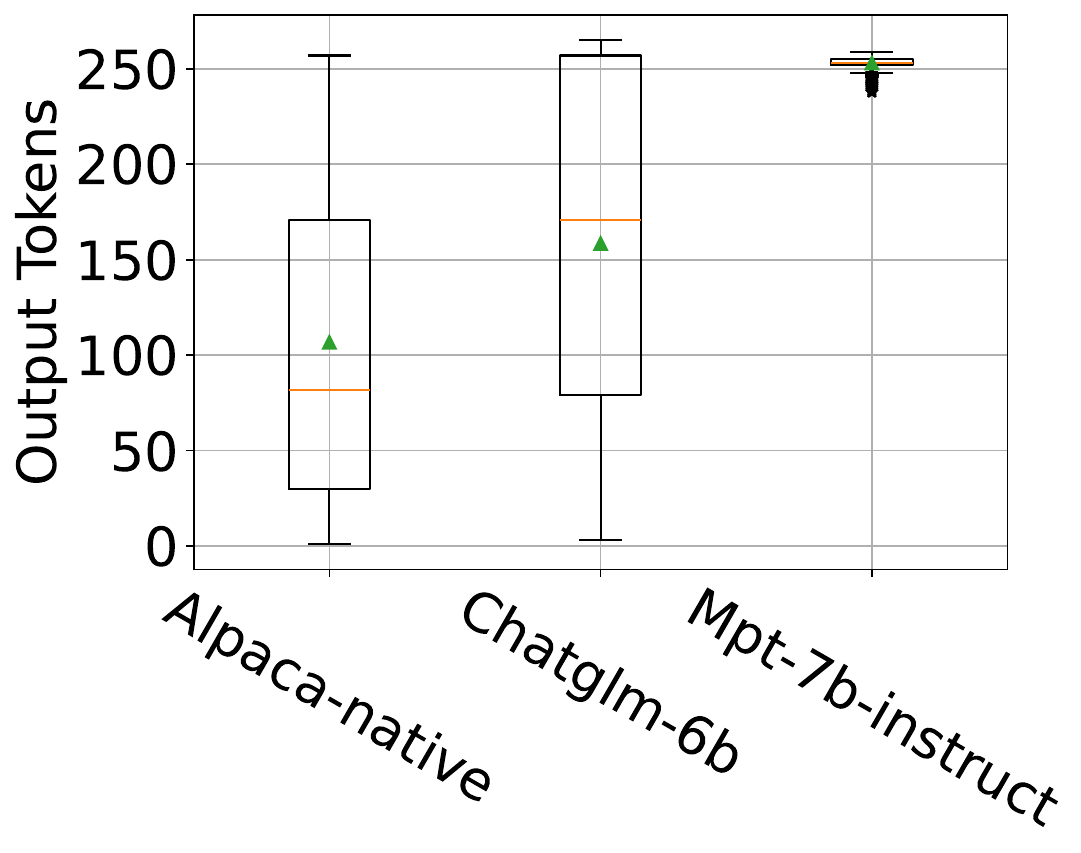} 
    }

    \caption{Distribution of response scores and response lengths across different LLMs.}
    \label{fig:pre2}
\end{figure}

Owing to being trained on diverse datasets, LLMs exhibit varying strengths and weaknesses when responding to different user requests. To intuitively demonstrate the heterogeneity of LLM services, we select 5,000 user requests from the mix-instruct dataset \cite{llm_blender} and conduct statistical analyses on the distribution of response quality and response length using Alpaca-native \cite{taori2023alpaca}, Chatglm-6b \cite{glm2024chatglm}, and Mpt-7b-instruct \cite{mosaicml2023introducing}, which are currently trending instruction-following LLMs. As depicted in Figure \ref{fig:pre2}, different LLMs exhibit distinct generative patterns. For instance, Mpt-7b-instruct has a slightly lower average response quality compared to the other two LLMs. Additionally, Mpt-7b-instruct tends to generate more tokens than the other two LLMs, with a more stable generation range. \textcolor{black}{Routing requests to Mpt-7b-instruct tends to consume more GPU memory while yielding relatively lower response quality. This further highlights the inherent heterogeneity among LLM services, which can lead to inefficient resource allocation and suboptimal user experience if not properly addressed.} Therefore, we should take into account the heterogeneity of LLM services while performing the LLM routing.

\subsection{Interference among Requests}
\label{subsec:interference}

\begin{figure}[t!]
    \centering
    \includegraphics[width=0.28\textwidth]{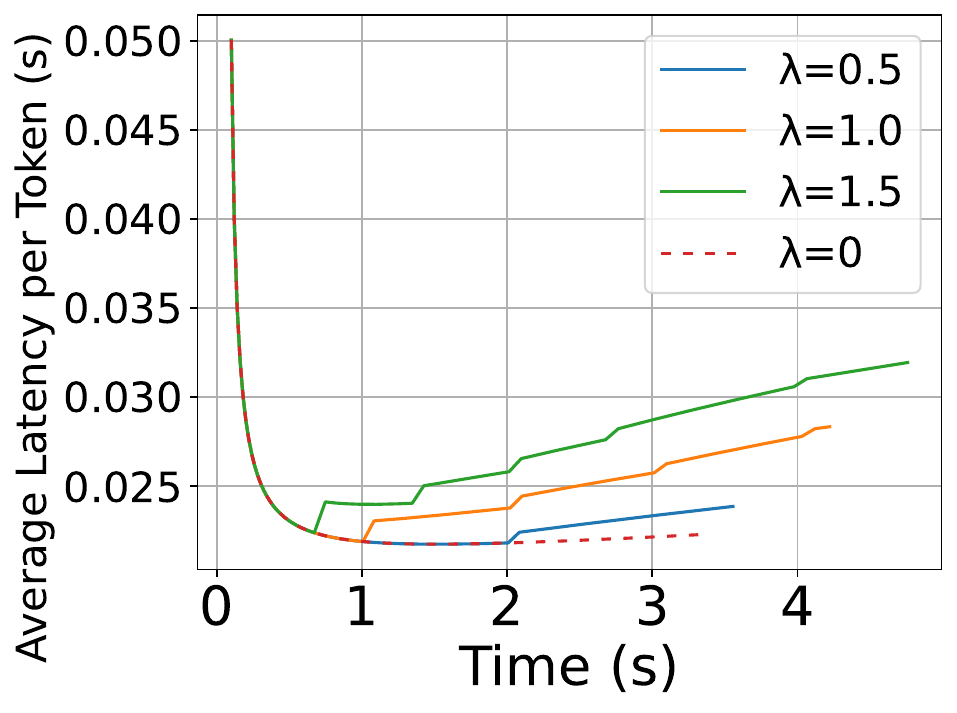} 
    \caption{Average latency per token over times for the first request with various request arrival rates.}
    \label{fig:interference}
\end{figure}

Continuous routing of new requests to a specific edge expert for processing can have implications for requests already queued. Specifically, during the iteration-level scheduling introduced in \cite{orca}, the prefill phase of newly enqueued requests can block the execution of running requests (e.g., the prefill phase of $q_2$ block the decoding process of $q_1$ as shown in Figure \ref{fig:inference}), while the decoding latency for these running requests increases due to the additional load from the new requests (e.g., the decode iteration 6 may be slightly slower than iteration 1 as shown in Figure \ref{fig:inference} due to an increase in the total number of tokens in the running queue). To investigate this phenomenon, we conduct experiments with various request arrival rates $\lambda$, tracking the average latency per token for the first request admitted to the queue over time, as depicted in Figure \ref{fig:interference}. The horizontal axis starts when the first request is enqueued and ends until completion.

From these experimental results, we observe the following: (i) There is a significant latency during the initial processing phase, specifically within the time it takes to generate the first token. This delay is primarily due to the prefill phase. (ii) After this initial latency spike, the latency per token stabilizes and then gradually increases due to the interference from incoming requests. (iii) As the request arrival rates $\lambda$ increase, the average latency per token rises more rapidly, indicating increased competition for computational resources. These observations highlight the impact of incoming requests on the average latency per token experienced by requests already in the queue, underscoring the critical need to account for request interference when making routing decisions.

\begin{table}[th!]
\renewcommand{\arraystretch}{1.3}
\caption{Main Notations.}
\label{table_example}
\centering
\begin{tabular}{cl}
\hline
\bfseries Notations & \bfseries Descriptions\\
\hline
$m_n$ & The $n$th edge expert.\\
$C_n$ & The GPU memory of edge expert $m_n$.\\
$t$ & The current time slot. \\
$Q_{n,t}^\text{waiting}$ & The waiting queue of edge expert $m_n$ at time slot $t$.\\
$Q_{n,t}^\text{running}$ & The running queue of edge expert $m_n$ at time slot $t$.\\
$q_j$ & The $j$th request.\\
$t_j$ & The arrival time slot of user request $q_j$.\\
$t_j^*$ & The completion time slot of user request $q_j$.\\
$p_j$ & The number of input tokens of request $q_j$.\\
$d_j$ & The total output length of request $q_j$.\\
$\hat{d}_{j}$ & The predicted output length of request $q_j$.\\
$d_{j,t}$ & The current output length of request $q_j$ at time $t$.\\
\multirow{2}{*}{$C_{j,n,t}$} & The required GPU memory of request $q_j$ processing on edge \\
& expert $m_n$ at time slot $t$.\\
$x_j$ & The routing decision of user request $q_j$.\\
$\hat{y}_j$ & The output response of user request $q_j$.\\
$y_j$ & The ground truth output response of user request $q_j$.\\
\multirow{2}{*}{$l_j$} & The final average latency per token of user request $q_j$ at the \\
& completion time slot $t_j^*$.\\
\multirow{2}{*}{$l_{j,t}$} & The current average latency per token of user request $q_j$ at \\
& time slot $t$. \\
\multirow{2}{*}{$l_{j,t}^+$} & The estimated increase in average latency per token of \\
& user request $q_j$ at time slot $t$. \\
\multirow{2}{*}{$\hat{l}_{j,t}$} & The estimated average latency per token of user \\
& request $q_j$ at time slot $t$. \\
$L$ & The system maximum latency requirement.\\
$s_j$ & The generation score of user request $q_j$.\\
$\hat{s}_j$ & The predicted generation score of user request $q_j$.\\
$\phi_j$ & The QoS for user request $q_j$.\\
\hline
\end{tabular}
\end{table}

\section{Problem Description}
\label{sec:problem}

In this section, we describe our scenario and formulate our QoS-aware LLM routing problem.

\subsection{Edge Computing with Multiple Experts Serving}
\label{subsec:edge_experts}

Today's edge servers are equipped with sufficient computational capabilities to support LLM inference services. As illustrated in Figure \ref{fig:infrastructure}, we consider a scenario where $N$ edge servers provide low-latency LLM services (i.e., edge LLM experts) to local edge devices. Given that these edge experts are LLMs specialized in different tasks, they possess distinct expertise. Moreover, these edge experts have varying computing resources, leading to significant heterogeneity. Each edge expert $m_n$ has a finite GPU memory $C_n$ dedicated to LLM inference tasks and processes user requests routed from the LLM router in the eAP. Specifically, at a given time slot $t$, each edge expert $m_n$ maintains a waiting queue $Q_{n,t}^\text{waiting}$ and a running queue $Q_{n,t}^\text{running}$ to efficiently manage multiple user requests. Upon routed from the LLM router at time slot $t_j$, user request $q_j$ is initially queued in $Q_{n,t_j}^\text{waiting}$ to await processing. To reduce queuing delays, techniques like iteration-level scheduling ~\cite{orca} are employed to manage running queue $Q_{n,t}^\text{running}$ and process requests. During each iteration at time slot $t$, user requests $q_j$ in $Q_{n,t}^\text{running}$ may occupy GPU memory resources $C_{j,n,t}$ across various decoding times due to the need to store intermediate results, activations, key-value caches, and other data. When request $q_j$ completes at time slot $t_j^*$, the user will receive the final output $\hat{y}_j$ from the edge expert, which consists of $d_j$ output tokens. Given that edge computing systems are located close to edge devices and offer substantial transmission bandwidth and our application scenario involves only the transmission of small volume text data, the impact on network quality is negligible.

\textcolor{black}{In the context of LLM services, the definition of QoS differs fundamentally from conventional services. Traditional services like web requests \cite{shen2023collaborative} and machine learning tasks \cite{crankshaw2017clipper,li2023tapfinger} (e.g., image classification and speech recognition) typically consider prediction accuracy as the response quality and end-to-end latency as the response latency. Considering that LLM services treat textual quality as a key aspect of response quality, we employ the BERTScore metric ~\cite{bertscore} to compare the output $\hat{y}_j$ of the edge expert against the ground truth text $y_j$, thereby quantifying the generation score $s_j = \text{BERTScore}(y_j, \hat{y}_j)$ for a given request $q_j$. Given the inherent token-by-token textual generation process of LLMs and the corresponding user reading pattern, we measure the average latency per token $l_j = (t_j^* - t_j) / d_j$ for request $q_j$ as the response latency. Additionally, we establish a maximum latency requirement $L$ that should be satisfied by all requests, meaning $l_j \leq L$. In practice, users generally have a positive experience when this condition is met; conversely, if the average latency per token $l_j$ exceeds latency requirement $L$, users might experience unacceptable delays, leading them to abandon or reissue their request. To formalize this, we define the QoS $\phi_j$ for a completed request $q_j$ as follows,
\begin{align}
    \label{equ:qos}
    \phi_j=s_j \times \mathbb{I}[l_j \leq L],
\end{align}
where the indicator function $\mathbb{I}[l_j \leq L]$ evaluates whether the request $q_j$ satisfies the predefined latency requirement $L$. This definition comprehensively considers both the response quality and response latency from the user's perspective, aligning with key QoS considerations in real-world LLM services.}

\subsection{Quality-of-Service Aware LLM Routing}
\label{subsec:multi_expert_Serving}
For simplicity, we define $[X] \triangleq \{1, 2,..., X\}$ in our description to represent the set of all integers from $1$ to $X$. When a request $q_j$ arrives at the eAP at time slot $t_j$, the LLM router within the eAP needs to make its routing decision $x_j \in [N]$ to route the request to one of the edge experts for processing. To make an optimal routing decision $x_j$, we formalize our QoS-aware LLM routing problem as a non-convex optimization problem. The objective of this optimization problem is to maximize the overall QoS $\phi_i$ of all requests $q_i$ while satisfying the GPU memory constraint of the edge experts. In practice, the QoS $\phi_i$ can only be assessed once the user request $q_i$ has been fully processed. Besides, the QoS $\phi_i$ of existing requests $q_i$ is determined by the edge expert handling them and can be affected by the routing decision $x_j$ of newly incoming request $q_j$ due to the potential interference among requests. Therefore, the QoS $\phi_i$ of each request $q_i$ is unknown as priori and can be affected by the routing decision $x_j$. Finally, we formulate the optimization problem as follows,
\begin{align}
    & \quad \mathop{\arg\max}\limits_{x_j} \quad \sum\limits_{i=1}^j \phi_i, \label{y} \\
    \text{s.t.} & \quad  x_j \in [N], \\
    & \sum\limits_{i \in Q_{n,t_j}^\text{running}} C_{i,n,t_j} \leq C_n, \quad \forall n \in [N],  \label{yy}
\end{align}
where Eq. (\ref{yy}) imposes the GPU memory constraint on each edge expert $m_n$ to ensure they can handle the processing load without exceeding their capacity $C_n$.

To solve this optimization problem, we face the following challenges (i) The ground truth generation score $s_i$, the ground truth output length $d_i$, and the final QoS $\phi_i$ can only be assessed once the user request $q_i$ has been fully processed. As a result, eAP cannot derive routing strategies by accurately solving the above optimization problem. (ii) Additionally, the pattern of user request arrivals is unknown, complicating the optimization process. (iii) Given the real-time nature of LLM routing, we need to efficiently solve this optimization problem once request $q_j$ arrives in the eAP while considering the optimization of long-term QoS. These challenges complicate the optimization problem such that traditional optimization methods are insufficient to address it.

\begin{figure*}[ht!]
\centering
\includegraphics[width=0.75\textwidth]{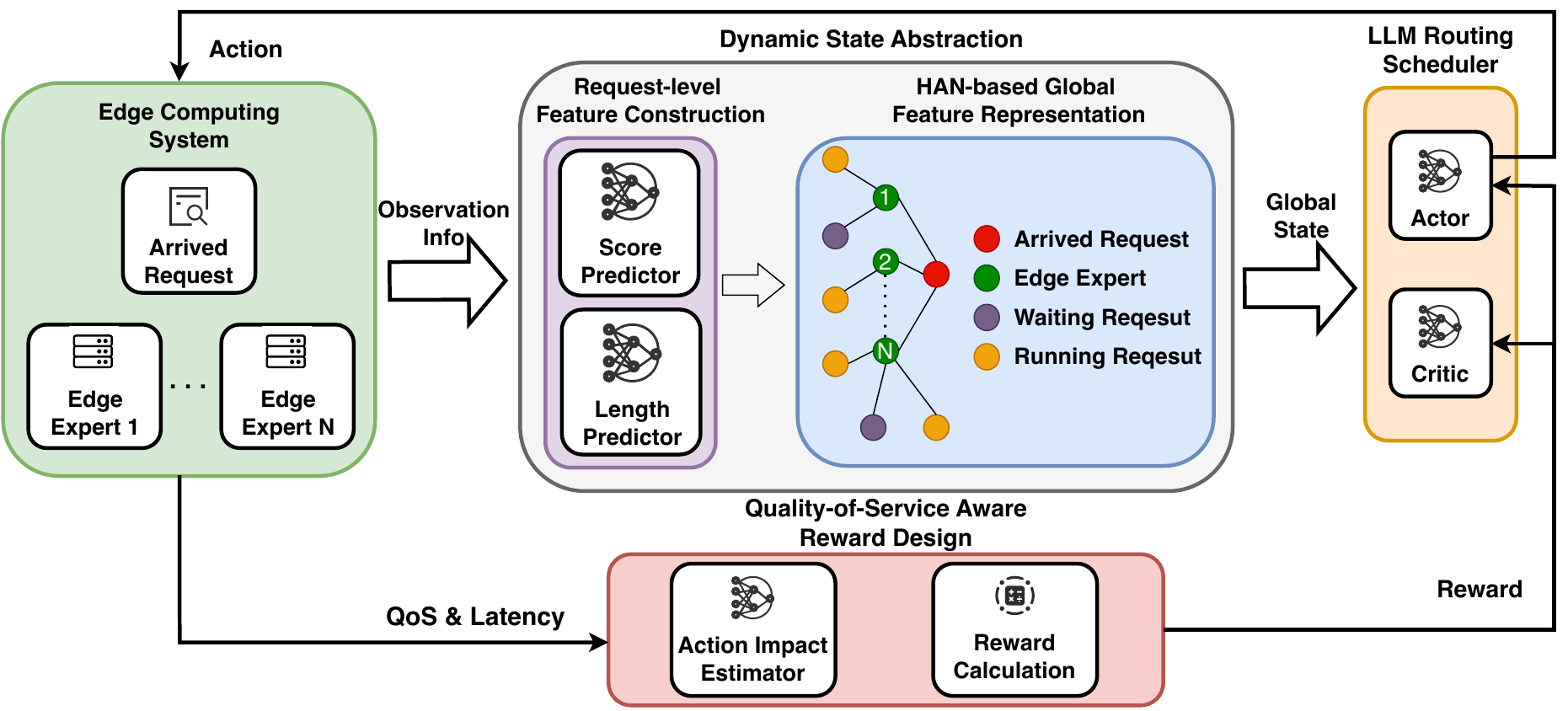}
\caption{Overview of our DRL-based QoS-aware LLM routing algorithm.}
\label{fig:overview}
\end{figure*}

\section{Algorithm Design}
\label{sec:algo}

To optimize the long-term QoS under dynamic workloads, we propose a DRL-based QoS-aware LLM routing algorithm, which leverages the adaptive learning and decision-making capabilities of DRL to efficiently handle varying workload conditions and optimize routing decisions. Figure \ref{fig:overview} illustrates the framework overview of our DRL-based QoS-aware LLM routing algorithm, highlighting the LLM routing algorithm based on the actor-critic architecture across multiple edge experts. To address the challenge of the dynamic nature of the system state, we propose a dynamic state abstraction technique based on heterogeneous graph attention network (HAN) to efficiently abstract dynamic system state features. Considering that the routing decisions can impact the overall QoS, we introduce an action impact estimator to assess the effects of routing decisions on overall QoS and design a QoS-aware reward based on this estimator.

\subsection{LLM Router by DRL}
\label{subsec:router}

Driven by the complexity of optimizing the routing decision variables in our optimization problem (\ref{y})-(\ref{yy}), traditional optimization methods fall short of providing effective solutions. DRL algorithms, however, excel at dealing with uncertainty and interacting with dynamic environments. They can learn the statistical patterns of resource sensitivities of requests, interference among requests, and heterogeneity of edge experts and optimize the long-term QoS $\sum\limits_{i=1}^j \phi_i$ in Eq. (\ref{y}) through trial and error. Additionally, the DRL agent can engage with various arrival pattern environments, which offers a significant advantage in handling dynamic workloads and generalizing to unseen workloads. Therefore, we propose utilizing the DRL algorithms to address the QoS-aware LLM routing problem. 

We formulate the QoS-aware LLM routing process as an infinite-horizon Markov Decision Process (MDP), defined by the tuple $(\mathcal{S}, \mathcal{A}, p, r, \gamma)$, where the state space $\mathcal{S}$ and the action space $\mathcal{A}$ are continuous, and the unknown state transition probability $p: \mathcal{S} \times \mathcal{A} \times \mathcal{S} \rightarrow [0, 1)$ represents the probability density of the next state $s_{j+1} \in \mathcal{S}$ given the current action $x_j \in \mathcal{A}$ and state $s_j \in \mathcal{S}$. The environment emits a bounded reward $r: \mathcal{S} \times \mathcal{A} \rightarrow [r_{\min}, r_{\max}]$ on each transition and $\gamma \in [0,\ 1)$ is the discount factor. Next, we brief the state space, action space, and reward in our DRL algorithm.

\textbf{State Space.} By monitoring the state of requests, the global state $s_j \in \mathcal{S}$ enables the LLM router to make more informed decisions. Due to the dynamic nature of the global state, we propose a dynamic state abstraction technique to abstract the global state features, introduced in Section \ref{subsec:state}.

\textbf{Action Space.} When request $q_j$ arrives at the eAP, the LLM router within eAP must decide whether to route request $q_j$ to one of the edge experts $m_n$ or to drop the request. Therefore, the action space is represented by an integer $x_j \in \{0, 1, ..., N\}$, where $0$ indicates dropping the request and $\{1, ..., N\}$ represent the available edge experts.

\textbf{Reward.} To guide the optimization of routing decisions, the reward function $r$ is considered to maximize the accumulated QoS while penalizing the incorrect routing decisions that negatively impact overall QoS. We design a QoS-aware reward introduced in Section \ref{subsec:reward}.

Due to the dynamic nature of workloads, DRL agents are prone to converging to suboptimal policies during the learning process. To achieve stable and rapid convergence to optimal policies, we adopt the Soft Actor-Critic (SAC) \cite{sac} algorithm for DRL training. By introducing entropy into the objective function, SAC encourages the agent to explore the state space more thoroughly, avoiding premature convergence to suboptimal policies. This approach promotes more diverse exploration and enhances the stability of the training process. The objective of our DRL agent is to find the policy $\pi(x_j|s_j)$, a distribution of actions over states, that maximizes the trade-off between the expected sum of rewards and the expected entropy defined as follows,
\begin{align}
    J(\pi) = \sum_{j=0}^\infty \mathbb{E}_{(s_j,x_j) \thicksim \rho_\pi} &\Big( \sum_{l=j}^\infty \gamma^{l-j} \mathbb{E}_{s_l\thicksim p,x_l\thicksim \pi} \big[r(s_j,x_j) \notag \\
    & + \alpha \mathcal{H}(\pi(\cdot|s_j)) | s_j,x_j \big] \Big),
\end{align}
where $\alpha$ is the temperature parameter that determines the relative importance of the entropy term against the reward, $\rho_\pi(s_j, x_j)$ denote the state-action marginals of the trajectory distribution induced by a policy $\pi(x_j|s_j)$ and $\mathcal{H}$ is the entropy function. This objective ensures that the policy not only achieves high rewards but also maintains exploration, leading to more robust and stable learning.

\subsection{Dynamic State Abstraction}
\label{subsec:state}
To characterize the global state, we first need to describe the state of each edge expert, including their computational resource utilization and workload conditions. However, the features of a request depend on its operational state and the edge expert handling it, both of which can vary and can influence the routing decisions. Therefore, we conduct a fine-grained characterization of the state for each request handled by each edge expert. Given the number of running and waiting requests at each edge expert is dynamic, simply stacking the global state features could result in a non-compact state space. To address this, we use a HAN to encode the features of different edge experts and the requests they manage, resulting in more compact environmental patterns and thereby enhancing the learning efficiency of the DRL agent.

\subsubsection{Request-level Feature Construction}
\label{subsec:request_state}

To represent the state of a request, we have meticulously defined its operational and expert-related features. The operational features capture the immediate output metrics of request, detailing the current output length $d_{j,t}$, the GPU memory utilization $e_{j,n,t} = C_{j,n,t} / C_n$ and the current average latency per token $l_{j,t} = (t - t_j) / d_{j,t}$. This provides a snapshot of the current operational efficiency for the DRL agent, enabling it to make routing decisions that take into account the request's current efficiency. The expert-related features, on the other hand, focus on the request features related to the edge expert handling the request, detailing the number of input tokens $p_j$, the future generation score $s_j$, and the total output length $d_j$. This provides a strategic perspective for the DRL agent, enabling it to factor in the expertise and resource capabilities of the handling edge expert when making routing decisions. Together, these features offer a comprehensive view of both the immediate metrics and the expert-related outcomes of the user requests, enabling more informed decision-making and process optimization. However, it is impossible to obtain the future generation score $s_j$ and the ground truth output length $d_j$ during the uncertain inference process of edge experts. Therefore, we train predictive models to assess the predicted generation score $\hat{s}_j$ and the predicted output length $\hat{d}_j$. These predictors take user request input text as input and predict the generation score and output length for user requests across different LLM services.

Since the instability of the generation process of edge experts, it is challenging to accurately predict the generation score $s_j$ and the total output length $d_j$ of each user request $q_j$ based on the request input text. Additionally, during the LLM routing decision process, we only need to roughly estimate a general range, such as an approximate generation score interval or output length interval, rather than exact values. Therefore, we bucketize the request generation scores and output lengths to predict their ranges. \textcolor{black}{This design intentionally incorporates tolerance for minor inaccuracies, allowing the DRL agent to learn robust routing patterns rather than overfitting to precise numerical dependencies.}

Specifically, we allocate the range of each bucket as $\frac{\text{max generation score}}{\text{number of buckets}}$ and $\frac{\text{max output length}}{\text{number of buckets}}$, respectively, and we properly use 10 buckets. DistilBERT \cite{distilbert} excels at processing and understanding text content, enabling it to provide accurate predictions for generation scores and output lengths. Moreover, DistilBERT's small size and fast inference speed make it highly suitable for real-time LLM routing scenarios. Therefore, we use these buckets as labels and request input text as inputs to fine-tune a DistilBERT model. Considering edge experts' heterogeneity introduced in Section \ref{sec:heter}, a trivial approach would be to train a specialized predictor for each edge expert, but this would incur a heavy computational overhead and storage overhead, which is not suitable for practical applications. Therefore, to reduce unnecessary overhead, we fine-tune only one model to predict each edge expert. To do this, we use a special token $<\text{extra\_token\_n}>$ to represent edge experts and add this special token before the request input text. Employing this method, we achieve a top-1 accuracy of 63.39\% in predicting generation score and 72.97\% in predicting output length. We also accomplish a top-3 accuracy of 97.78\% in predicting the generation score and 84.71\% in predicting the output length. Additionally, we measure that our predictors take 5ms to run on an NVIDIA RTX 4090 GPU, which is negligible compared to the total request processing time.

To this end, given a user request $q_j$ processed in edge expert $m_n$ at time slot $t$, the user request state $\boldsymbol{f}_{q_j, t}$ is defined as follows,
\begin{align}
    \boldsymbol{f}_{q_j, t} = (p_j,\ \hat{s}_j,\ \hat{d}_j,\ e_{j,n,t},\ d_{j,t},\ l_{j,t}),
\end{align}
where $\hat{s}_j$ is the predicted generation score of request $q_j$ and $\hat{d}_j$ is the predicted output length of request $q_j$. For different types of requests, whether they are running, waiting, or under routing, their state can be represented by characterizing these two categories of features mentioned above.

\subsubsection{HAN-based Global State Representation}
\label{subsec:global_state}

In addressing our QoS-aware LLM routing problem, an effective router should consider both the workload on each edge expert and their GPU memory utilization. It should also adapt to GPU memory utilization of different requests which affect the QoS. A key insight is that the running requests can continuously provide GPU memory utilization information, and the waiting requests indicate their increased latency and future contention. Therefore, the global state needs to constantly monitor running requests, waiting requests, and their GPU memory utilization.

Given the need to constantly monitor running requests, waiting requests, and their GPU memory utilization, we define the edge expert state $\boldsymbol{f}_{m_n, t}$ as follows,
\begin{align}
    \boldsymbol{f}_{m_n, t} = (e_{n,t},\ &|Q_{n,t}^\text{running}|,\ |Q_{n,t}^\text{waiting}|,\ \notag \\ 
    & \{\boldsymbol{f}_{q_j, t},\ \forall j \in Q_{n,t}^\text{running} \cup Q_{n,t}^\text{waiting}\}),
\end{align}
where consists of the total GPU memory utilization $e_{n,t}  = \sum\limits_{j \in Q_{n,t}^\text{running}} C_{j,n,t} / C_n$, the number of running requests $|Q_{n,t}^\text{running}|$, the number of waiting requests $|Q_{n,t}^\text{waiting}|$ and all the state features of requests $q_j$ in the running queue $Q_{n,t}^\text{running}$ and waiting queue $Q_{n,t}^\text{waiting}$. Notice that the final term of edge expert state $\boldsymbol{f}_{m_n, t}$ is time-varying. For instance, running requests will dequeue and free the occupied GPU memory when completed. If sufficient GPU memory becomes available, the first waiting request will move to the running queue, while new requests will initially be placed in the waiting queue. Therefore, not only does the state of the requests change, but the number of running and waiting requests is also dynamic. Finally, we define the raw global state features $\boldsymbol{f}_t$ as follows,
\begin{align}
    \boldsymbol{f}_t = \{\boldsymbol{f}_{q_i, t},\ \boldsymbol{f}_{m_1, t},\ ...,\ \boldsymbol{f}_{m_n, t}\},
\end{align}
where consists of the state features $\boldsymbol{f}_{q_i, t}$ of arrived request $q_i$ and all the state features of edge experts.

Since the dynamic nature of request arrivals, the final term of edge expert state $\boldsymbol{f}_{m_n, t}$ is time-varying. To provide a fixed-size global state for the DRL agent, we need to pad the running and waiting queue of each edge expert to a sufficiently large size based on the system workload. In this case, we cannot simply stack $\boldsymbol{f}_t$ into a global state matrix and feed it to the DRL agent. The main drawbacks are: (i) state space is not compact due to the redundant padding features. (ii) the graph structure and semantic relation between requests and edge experts will be lost. Instead, our solution embeds the entire graph into a neural network and enables iterative state interaction across edge experts.

Due to the dynamic nature and the semantic graph structure of the global state, we adopt a HAN to embed global state features due to its ability to effectively capture heterogeneous graph information and adaptively prioritize significant relationships, overcoming the limitations of traditional GNNs in handling complex and dynamic semantic graph structures. As shown in Figure \ref{fig:overview}, we collect the global state features $\boldsymbol{f}_t$ and construct our heterogeneous graph $G_t(V_t, E_t)$, where the node set $V_t$ consists of arrived request node, edge expert nodes, running request nodes and waiting request nodes. The edges in $E_t$ are defined as follows. Each running request node connects with the edge expert node on which it is executing. Analogously, each waiting request node connects with the edge expert node it is waiting in. Finally, an arrived request node connects with all edge expert nodes. The information propagation of HAN passes the features as messages from the neighbors to each node $u \in V_t$ and aggregates them with the features of $u$ using a two-level attention network in a configurable number of interactions. The propagation model of our HAN is formalized as follows,
\begin{align}
    & G_t^{j, (0)} = \boldsymbol{f}_{q_j, t},   \label{x} \\
    & G_t^{n, (0)} = (e_{n,t},\ |Q_{n,t}^\text{running}|,\ |Q_{n,t}^\text{waiting}|),\quad \forall n \in [N], \\
    & G_t^{i, (0)} = \boldsymbol{f}_{q_i, t},\quad \forall i \in \bigcup\limits_{n \in [N]} Q_{n, t}^\text{running}, \\
    & G_t^{k, (0)} = \boldsymbol{f}_{q_k, t},\quad \forall k \in \bigcup\limits_{n \in [N]} Q_{n,t}^\text{waiting}. \label{xx}
\end{align}

We denote the initial global state input as $G_t^{(0)} = \{G_t^{j, (0)},\ G_t^{n, (0)},\ G_t^{i, (0)},\ G_t^{k, (0)} \}$, as in (\ref{x})-(\ref{xx}). The node embedding is propagated in each layer $l$, i.e., $G_t^{(l)} = g(G_t^{(l-1)})$, where $g(\cdot)$ represents the two-level attention network aggregating the features of each node with its neighbors. After $L$ layers of graph message passing, we get the final graph embedding $G_t^{(L-1)}$. We then map the arrived request node embedding $G_t^{j,(L-1)}$ as the input of the DRL agent.

\subsection{Quality-of-Service Aware Reward Design}
\label{subsec:reward}

As introduced in Section \ref{subsec:interference}, routing decisions can influence the average latency per token experienced by requests already in the queue, thereby affecting the overall QoS. To evaluate the effects of routing decisions on overall QoS, we propose an action impact estimator that estimates the prefill and decode latencies for incoming requests and analyzes the impact of request interference within an edge expert on overall QoS. Building on this estimator, we design the QoS-aware reward, which penalizes the negative effects of routing decisions on overall QoS.

\subsubsection{Action Impact Estimator}
\label{subsec:estimate}
Given that the latency is primarily caused by the prefill and decode phases in LLM inference, to estimate the impact of routing decisions on overall QoS, we first estimate the prefill and decode latencies for the incoming requests. During iteration-level scheduling, the latency of the incoming request during the prefill phase increases rapidly and linearly with an increase in the number of input tokens. Conversely, the decode phase has a minimal impact, with the mean decode time increasing slowly as the total tokens grow. Thus, we estimate the prefill latency $l_{j,t}^{\text{pre}}$ and decoding latency $l_{j,t}^{\text{dec}}$ for request $q_j$ when batch executed in edge expert $m_n$ at time slot $t$ as follows,
\begin{align}
    & l_{j,t}^{\text{pre}} = k_{1,n} \times p_j, \\
    & l_{j,t}^{\text{dec}} = k_{2,n} \times \sum\limits_{i \in Q_{n,t}^{\text{running}}} (p_i + d_{i,t}),
\end{align}
where $k_{1,n}$ and $k_{2,n}$ represent the gradient of prefill phase and decode phase, respectively, determined through profiling of edge expert $m_n$. 

Batching requests in an edge expert can impact the overall QoS in two primary ways: (i) the prefill phase of the incoming requests will block the running requests, and (ii) the decoding latency for these running requests increases due to the additional load from new requests. Suppose request $q_j$ is routed to edge expert $m_n$ (i.e. routing decision $x_j = n$) at time slot $t_j$. Based on the aforementioned analysis, we estimate the increase in average latency per token $l_{i,t_j}^+$ for all requests $q_i \in Q_{n,t_j}^{\text{running}}$ due to the incoming request $q_j$ as follows,
\begin{equation}
    l_{i,t_j}^+ = \frac{1}{d_i} ( k_{1,n} \times p_j + k_{2,n} \times \sum\limits_{k=1}^{\min(d_i-d_{i,t_j},\ d_j)} (p_j + k)),
\end{equation}
where the first term indicates the increased latency caused by the prefill phase of incoming request $q_j$ and the second term indicates the increased decoding latency caused by the additional load from request $q_j$. Subsequently, the estimated average latency per token for all requests $q_i \in Q_{n,t_j}^{\text{running}}$ is calculated as $\hat{l}_{i,t_j} = l_{i,t_j} + l_{i,t_j}^+$. The impact of the routing decision $x_j$ on overall QoS is then given by $\sum\limits_{i \in Q_{n,t_j}^{\text{running}}} \phi_i \times \mathbb{I}[\hat{l}_{i,t_j} \geq L]$, where the indicator $\mathbb{I}[\hat{l}_{i,t_j} \geq L]$ determines whether request $q_i$ will exceed the latency requirement $L$.

\subsubsection{Quality-of-Service Aware Reward}
The design of the reward function must take into account both the accumulated QoS for all requests that meet the latency requirement and the interference among requests within the selected edge expert $m_{x_j}$ resulting from the current routing action $x_j$. The arrival of new requests can affect the average latency per token of other running requests within the same edge expert, potentially causing some to exceed the latency requirement. Such impacts should be penalized to prevent latency requirement violations and maximize the overall QoS. Building on the analysis presented in Section \ref{subsec:estimate}, the reward $r_{j}$ after making a routing action $x_j$ at time slot $t_j$ is defined as follows,
\begin{equation}
\begin{split}
    r_{j} =  \sum\limits_{n=1}^N \sum\limits_{i \in Q_{n,t_j}^{\text{running}}} & \phi_i \times w_{n,i,t_j} \times \mathbb{I}[l_i \leq L] \\
    & - \sum\limits_{i \in Q_{x_j,t_j}^{\text{running}}} \phi_i \times \mathbb{I}[\hat{l}_{i,t_j} \geq L],
\end{split}
\end{equation}
where $w_{n,i,t_j} \in \{0,1\}$ indicates whether the request $q_i$ was completed by edge expert $m_n$ at time slot $t_j$ and $Q_{x_j,t_j}^{\text{running}}$ indicates the running queue of the selected edge expert $m_{x_j}$ at time slot $t_j$. The first term represents a positive reward for each completed request that meets the latency requirement, reflecting the accumulated QoS achieved. The second term is a penalty for the estimated negative impact on overall QoS, assessing the potential adverse effects of the current routing decision $x_j$, thereby preventing violations of the latency requirement. During the training process, the training environment emits the reward $r_j$ after the DRL agent makes a routing decision $x_j$. The DRL agent will use this reward to refine its routing strategy, enhancing its decision-making capabilities through continuous trial and error. Upon successful completion of training, the trained DRL agent will be deployed on the eAP for practical LLM routing.

\subsection{Computational Efficiency of QoS-aware Router}
\label{sec:comp_analy}

\begin{table}[htb] 
\centering 
\caption{Component-wise Computational Profile of QoS-aware Router.}
\label{tab:computation_complexity} 
\begin{tabular}{c|c|c} 
\hline \textbf{Component} & \textbf{Parameter} & \textbf{Latency} \\ 
\hline Generation Score Predictor & 67M & 5ms  \\ 
Output Length Predictor & 67M & 5ms  \\ 
HAN & 19K & $<$ 1ms \\ 
Actor-Critic & 10K & $<$ 1ms \\ \hline
\end{tabular} 
\end{table}

\textcolor{black}{To quantify the computational overhead, we conduct comprehensive profiling of both model size and inference latency across all components. As shown in Table ~\ref{tab:computation_complexity}, the integrated components of our QoS-aware router, including the dynamic state abstraction model implemented via HAN and DistilBERT predictors, as well as the DRL actor-critic architecture, contain merely 134M model parameters, which is remarkably fewer than the billions of parameters of edge expert models. This makes our QoS-aware router highly computationally efficient while maintaining extremely low inference latency. Benefiting from parallel computation across independent predictors, empirical measurements show that our QoS-aware router requires only 5ms on an NVIDIA RTX 4090 GPU which is negligible compared to the multi-second generation latency of edge experts. Consequently, our QoS-aware router is sufficiently lightweight and computationally efficient, making it ideally suited for resource-constrained edge deployments. Furthermore, considering the rapid advancement in edge computing capabilities \cite{li2023tapfinger}, our lightweight QoS-aware router is well-suited for real-time routing scenarios with strict latency requirements.}

\section{Evaluation}
\label{sec:evaluation}

\subsection{Experiment Settings}
\label{sec:settings}

\textbf{Model Configurations.} For the implementation, we develop our DRL-based QoS-aware LLM routing algorithm using PyTorch \cite{pytorch} and leverage TorchRL \cite{torchrl}, a reinforcement learning library built on PyTorch, to manage the model training process. To implement the HAN, we utilize the PyTorch Geometric library \cite{fey2019fast} to accelerate the data loading, training, and inference efficiency. Our HAN configuration includes 2 layers with 4 attention heads to produce embeddings with a hidden size of 64. By default, the capacity of the running queue and waiting queue for each edge expert is set to 5. The routing action is determined by a two-layer perceptron. As for the critic, we employ a two-layer perceptron that takes the HAN embedding of the arrived request node as its input. To ensure the stable training of the DRL agent and to facilitate fast convergence, we employ the SAC algorithm \cite{sac} for training our DRL agent. Additionally, we conduct training over 1 million steps and save the models that achieve the best evaluation results.

\textbf{Baseline.} To validate the effectiveness of our proposed algorithm, we consider two heuristics and two representative algorithms as our baselines.
\begin{itemize}
    \item \textbf{BERT Router (BR).} Most existing works \cite{stripelis2024polyrouter, hu2024routerbench, zooter} use a fully fine-tuned BERT model \cite{devlin-etal-2019-bert} for LLM routing. To adapt it to our problem, we append a classification head with a softmax activation function on top of the BERT model and use the BERTScore as the label for training. This model chooses the edge expert with the highest predicted BERTScore.
    \item \textbf{Round-Robin (RR).} This method sequentially assigns each incoming user request to an edge expert, a common approach in web applications for load balancing.
    \item \textbf{Shortest Queue First (SQF).} This method prioritizes edge experts with the fewest requests in their queue, selecting the edge expert with the shortest queue to balance workload and reduce overall latency.
    \item \textbf{Baseline RL.} Existing DRL-based request routing algorithms \cite{jain2024intelligent} are designed across homogeneous LLM instances. They generally use expert-level features, such as expert resource utilization and queue situation, as raw state features. Besides, their reward designs are not suitable for our QoS-aware LLM routing scenario. To adapt these algorithms for our scenario and to effectively compare the performance of our design, we propose a modified RL algorithm that omits dynamic state abstraction and QoS-aware reward. This baseline algorithm employs raw expert-level features without dynamic state abstraction,
    \begin{equation*}
        \boldsymbol{f}_t = \{\boldsymbol{f}_{m_1, t},\ ...,\ \boldsymbol{f}_{m_n, t}\},
    \end{equation*}
    where $\boldsymbol{f}_{m_n, t} = (e_{n,t},\ |Q_{n,t}^\text{running}|,\ |Q_{n,t}^\text{waiting}|)$ are features of each expert $m_n$. Additionally, the reward function is formulated as follows,
    \begin{equation*}
        r_j = \sum\limits_{n=1}^N \sum\limits_{i \in Q_{n,t_j}^{\text{running}}}  \phi_i \times w_{n,i,t_j} \ , 
    \end{equation*}
    where $w_{n,i,t_j} \in \{0,1\}$ indicates whether the request $q_i$ completed at time $t_j$ by edge expert $m_n$.
\end{itemize}

\textbf{Environment Simulation.} To simulate user request content, we utilize the mix-instruct dataset \cite{llm_blender}, which comprises responses from currently trending instruction-following LLMs along with their corresponding evaluations. For our edge experts, we select up to 12 LLMs, each with approximately 7B parameters, and use BERTscore \cite{bertscore} for evaluating the quality of generated responses. Besides, We allocate a dedicated bandwidth of 1 Mbps for each connection between the eAP and edge experts. Our experiments are conducted under Poisson-distributed workloads and long-term real-world workloads, characterized by the request arrival rate $\lambda$. By default, the latency requirement $L$ for user requests is set to 30 milliseconds. Each edge expert is equipped with an NVIDIA RTX 4090 GPU and is deployed using vLLM \cite{vllm}, a state-of-the-art inference serving system.

\begin{figure}[t!]
    \centering
    \subfloat[Average QoS]{
		\includegraphics[width=0.23\textwidth]{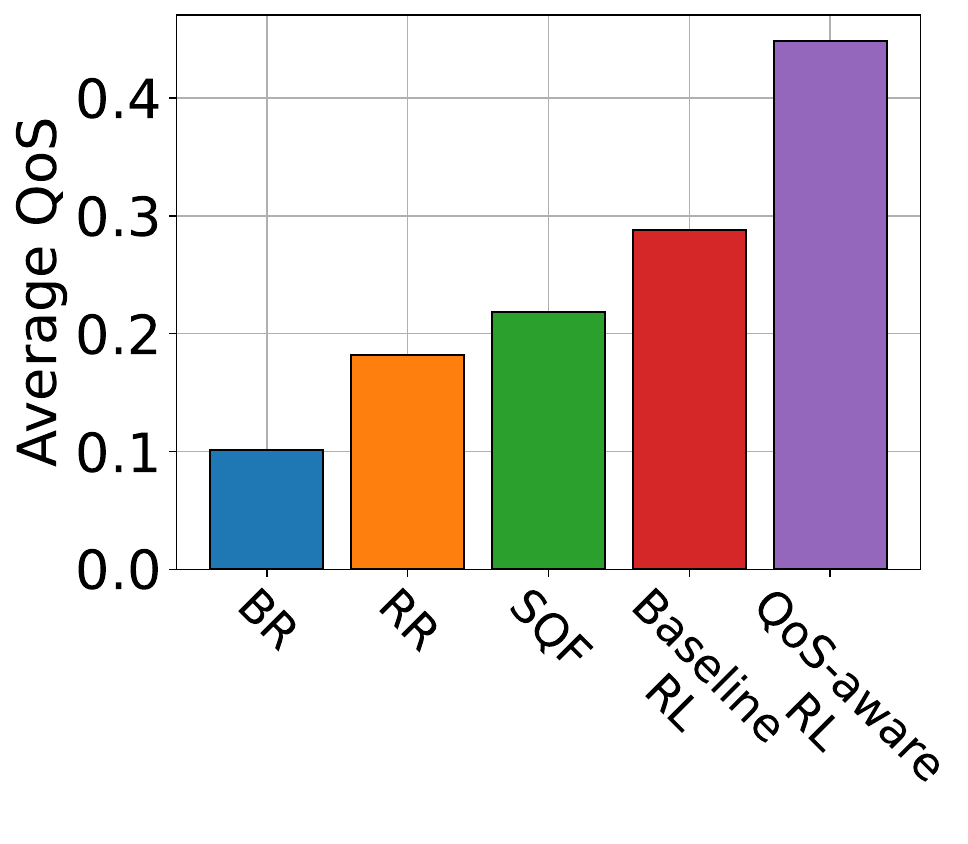}
		\label{fig:qos_poisson_5.0_6}
    }
    \hfil
    \subfloat[Average latency per token]{
		\includegraphics[width=0.23\textwidth]{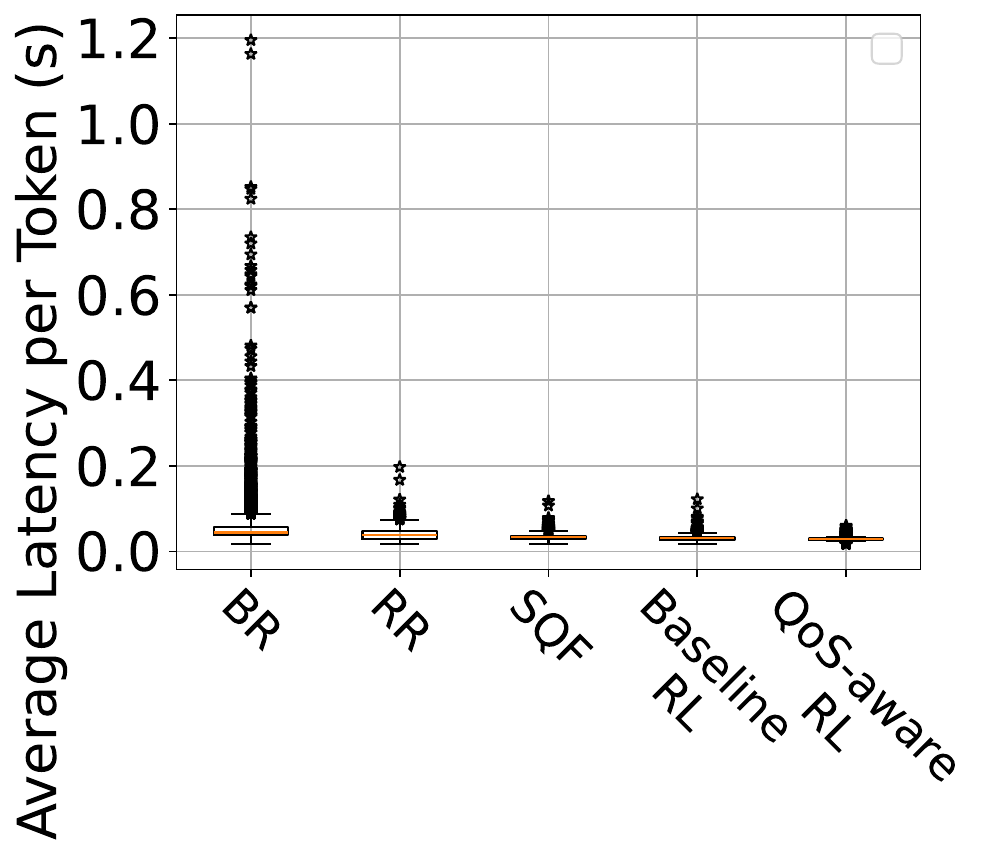} 
		\label{fig:latency_poisson_5.0_6}
    }
    \caption{Average QoS and average latency per token comparison with $N$=6 edge experts under Poisson workloads with $\lambda$=5.}
    \label{exp:main_poisson}
\end{figure}

\begin{figure}[t!]
    \centering
    \includegraphics[width=0.40\textwidth]{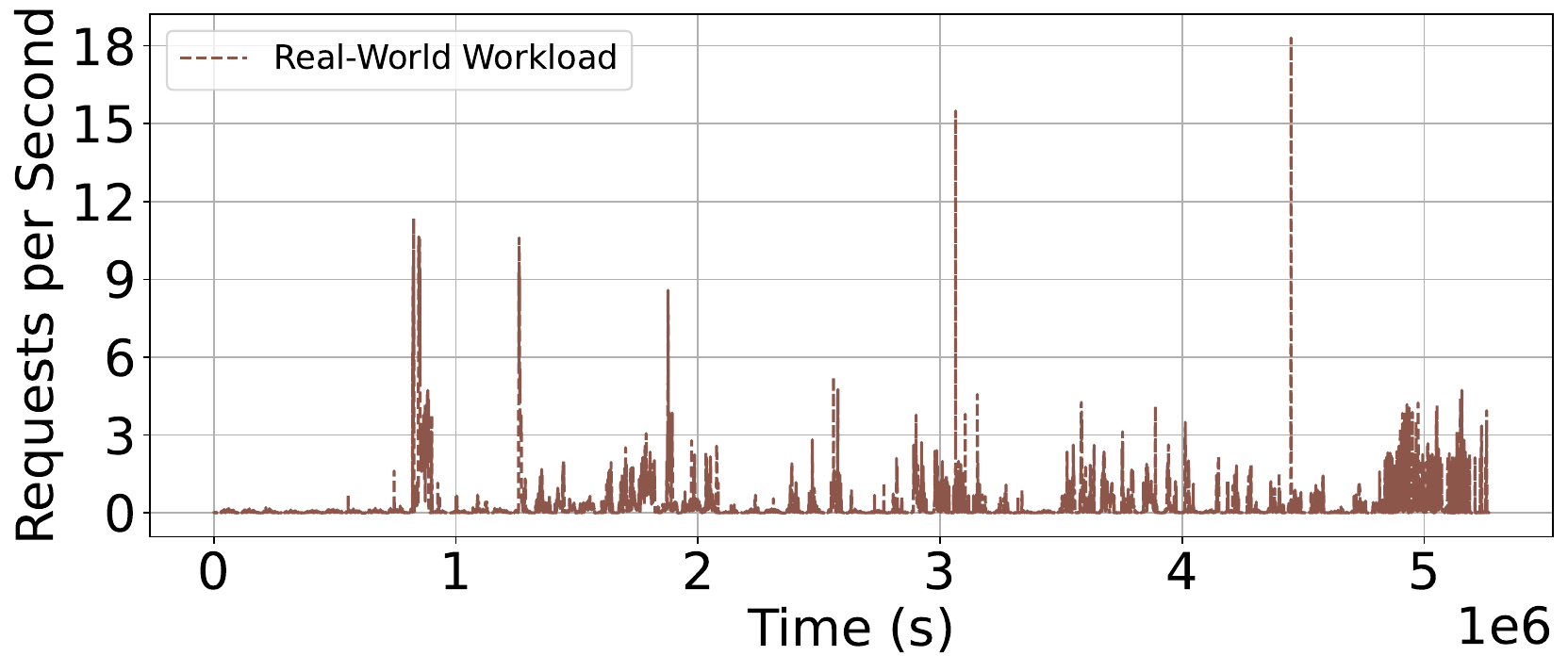}
	\label{fig:real_world_workload}
    \caption{Illustration of dynamic intensity under real-world LLM workloads.}
\end{figure}

\begin{figure}[t!]
    \centering
    \subfloat[Average QoS]{
		\includegraphics[width=0.23\textwidth]{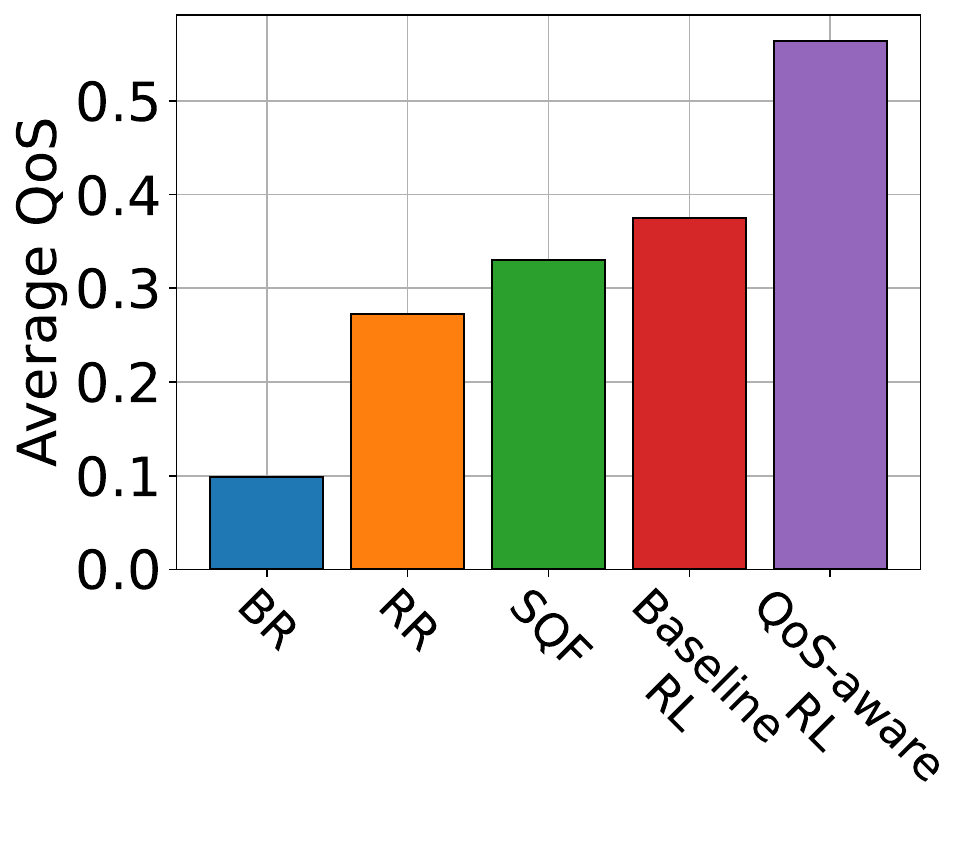}
		\label{fig:qos_real_4.5_6}
    }
    \hfil
    \subfloat[Average latency per token]{
		\includegraphics[width=0.23\textwidth]{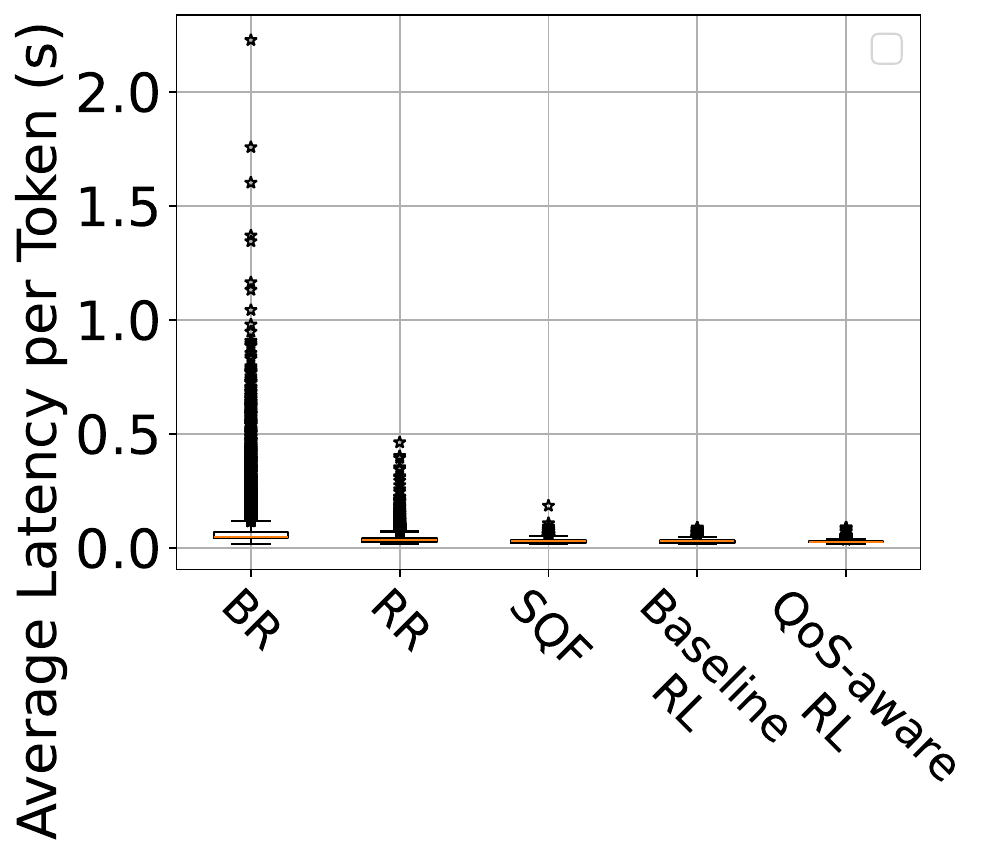} 
		\label{fig:latency_real_4.5_6}
    }
    \caption{Average QoS and average latency per token comparison with $N$=6 edge experts under long-term real-world LLM workloads.}
    \label{exp:main_real}
\end{figure}

\subsection{Performance}
\label{subsec:performance}

\paragraph{Evaluation on Poisson Workloads} As shown in Figure \ref{exp:main_poisson}, we present a comprehensive analysis of the average QoS and the average latency per token compared to baselines with $N$=6 edge experts under Poisson workloads with $\lambda$=5. The BERT-Router persistently routes requests to edge experts with high predicted generation scores while disregarding dynamic workloads, leading to inferior performance. In contrast, the Round-Robin and Shortest Queue First methods account for queue conditions but fail to incorporate QoS for user requests, resulting in poor performance. The Baseline RL approach surpasses other baselines due to its consideration of expert-level operation state and overall QoS for user requests. Meanwhile, our proposed algorithm further enhances performance through dynamic state abstraction, which captures fine-grained request-level features, and QoS-aware reward that considers the impact of each routing decision on overall QoS. Compared to Baseline RL, our proposed algorithm achieves a 35.78\% improvement in the average QoS and a 5.45\% reduction in the average latency per token.

\paragraph{Evaluation on Real-world Workloads} Our router is trained with $N$=6 edge experts under Poisson workloads with $\lambda$=5. To evaluate our algorithms under longer and more volatile workloads than that in the training stage, we further conduct several long-term experiments on real-world LLM service workloads provided by BurstGPT \cite{burstgpt}. As illustrated in Figure \ref{fig:real_world_workload}, the long-term real-world LLM workloads are challenging due to the dynamic request arrival intensity of user requests. Specifically, we select a period during which the average request arrival rates $\lambda=5$ for evaluation. Figure \ref{exp:main_real} demonstrates that our proposed algorithm maintains scalable performance even under highly volatile real-world LLM workloads. Compared to baseline methods, our proposed algorithm achieves at least a 33.47\% improvement in the average QoS while reducing the average latency per token by 3.35\%.

\begin{figure}[t!]
    \centering
    \includegraphics[width=0.40\textwidth]{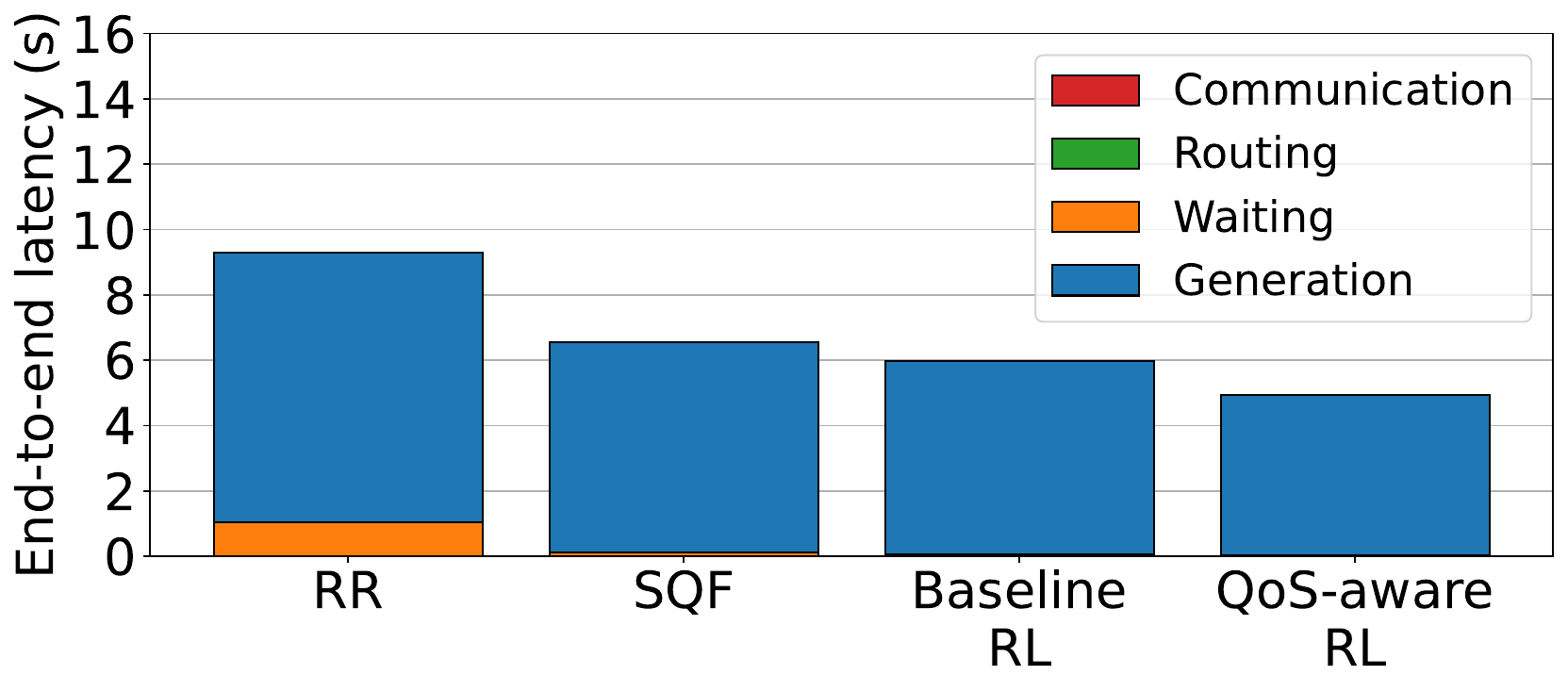}
    \label{fig:e2e_latency}

    \caption{Comparison of end-to-end latency with $N$=6 edge experts under Poisson workloads with $\lambda$=5.}
    \label{exp:e2e_latency}
\end{figure}

\begin{figure}[t!]
    \centering
    \subfloat[Average QoS]{
		\includegraphics[width=0.23\textwidth]{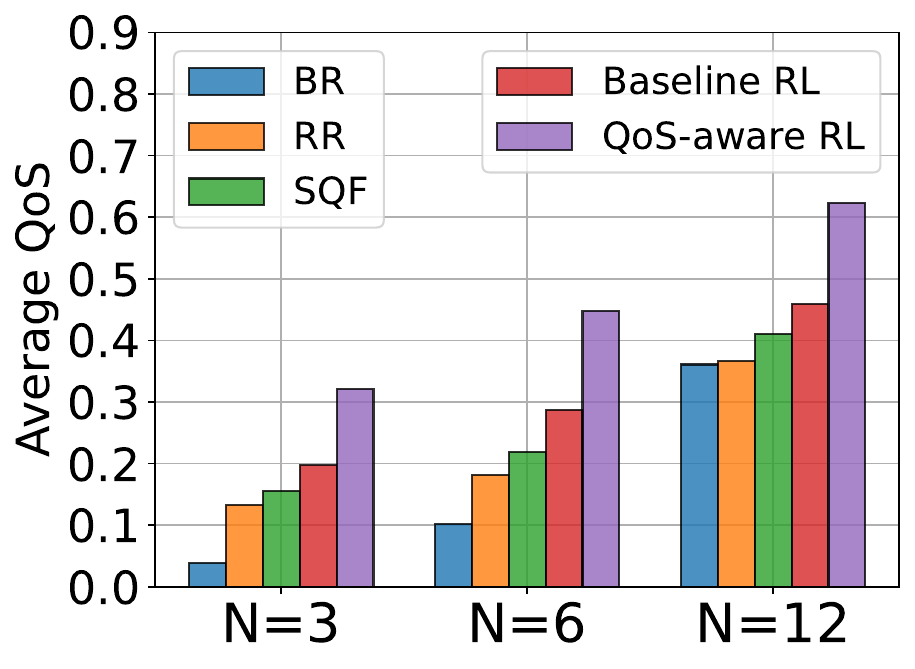}
		\label{fig:qos_poisson_[3,6,12]}
    }
    \hfil
    \subfloat[Average latency per token]{
		\includegraphics[width=0.23\textwidth]{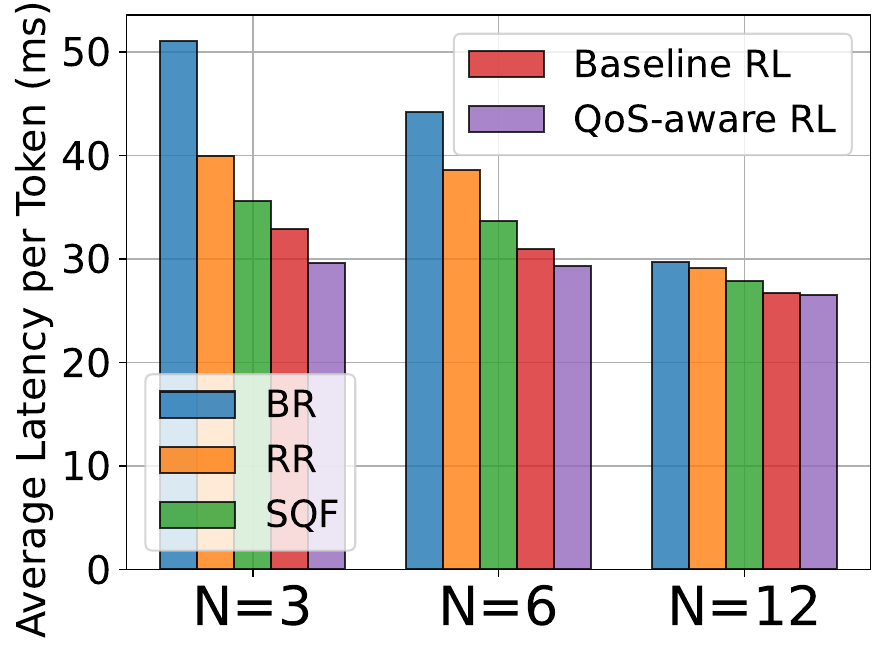} 
		\label{fig:latency_poisson_[3,6,12]}
    }

    \caption{Comparison of average QoS and average latency per token across increasing numbers of edge experts $N$ under Poisson workloads with $\lambda$=5.}
    \label{exp:varying_N}
\end{figure}

\begin{figure}[t!]
    \centering
    \subfloat[Average QoS]{
		\includegraphics[width=0.23\textwidth]{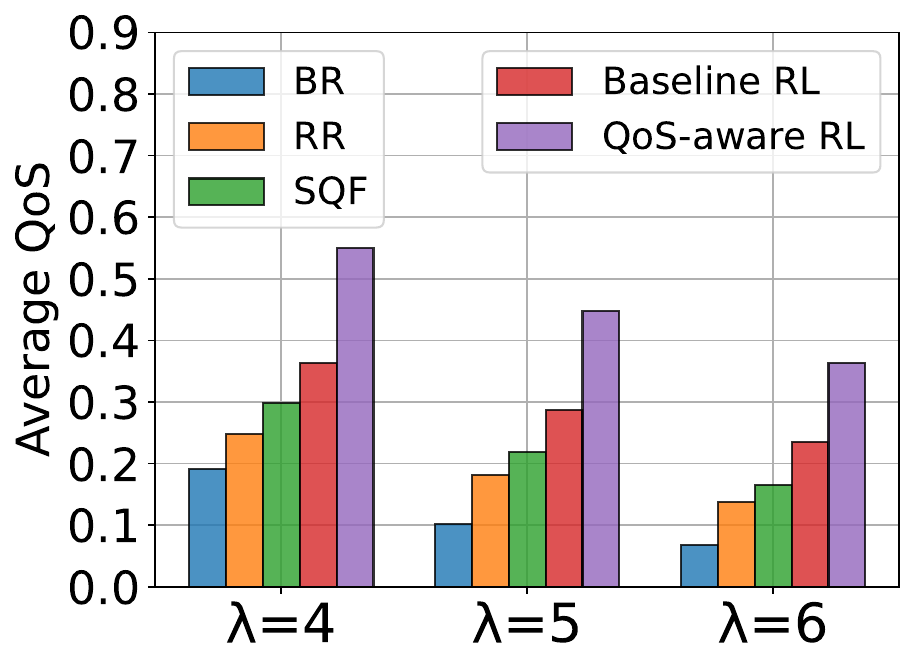}
		\label{fig:qos_poisson_[4.0,5.0,6.0]_6}
    }
    \hfil
    \subfloat[Average latency per token]{
		\includegraphics[width=0.23\textwidth]{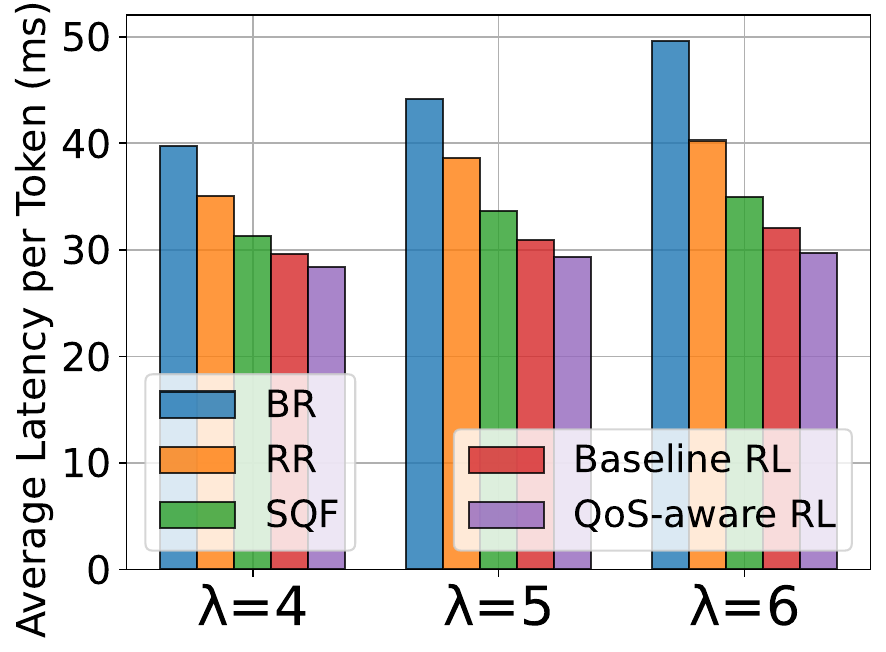} 
		\label{fig:latency_poisson_[4,5,6]_6}
    }

    \caption{Comparison of average QoS and average latency per token with $N$=6 edge experts across Poisson workloads with varying $\lambda$.}
    \label{exp:varying_lam}
\end{figure}

\begin{figure}[t!]
    \centering
    \subfloat[Average QoS]{
		\includegraphics[width=0.23\textwidth]{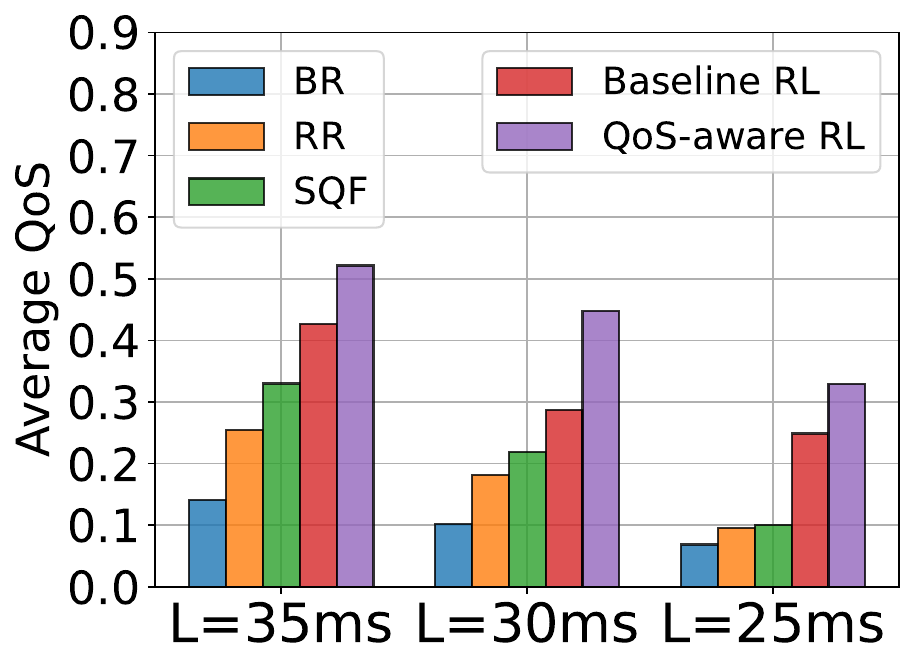}
		\label{fig:qos_poisson_[0.035,0.03,0.025]_6}
    }
    \hfil
    \subfloat[Average latency per token]{
		\includegraphics[width=0.23\textwidth]{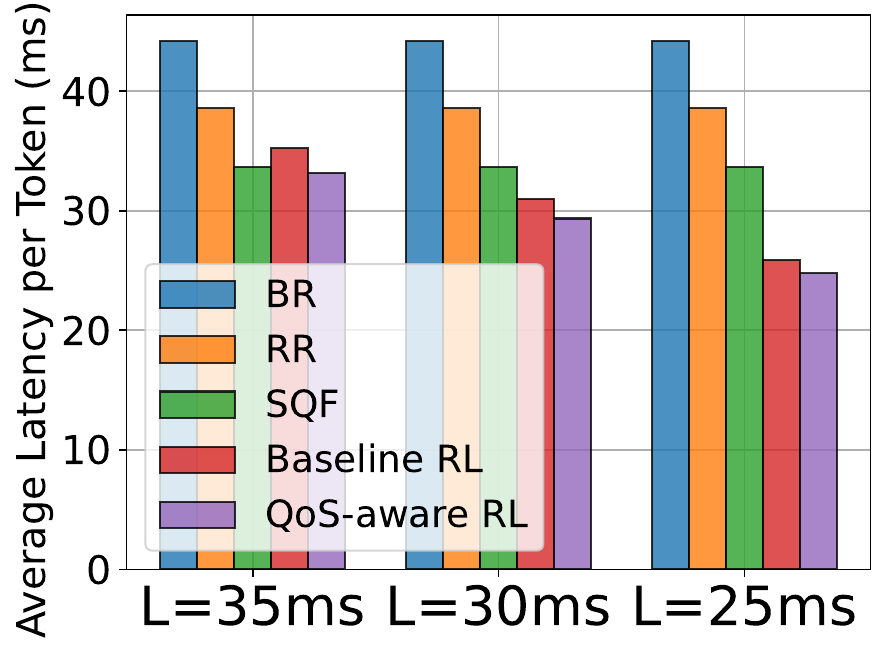} 
		\label{fig:latency_poisson_[0.035,0.03,0.025]_6}
    }
    
    \caption{Comparison of average QoS and average latency per token with $N$=6 edge experts under Poisson workloads with $\lambda$=5 across varying latency requirements $L$.}
    \label{exp:varying_L}
\end{figure}

\subsection{Analysis}

\begin{figure*}[t!]
    \centering
    \subfloat[BR vs QoS-aware RL]{
		\includegraphics[width=0.23\textwidth]{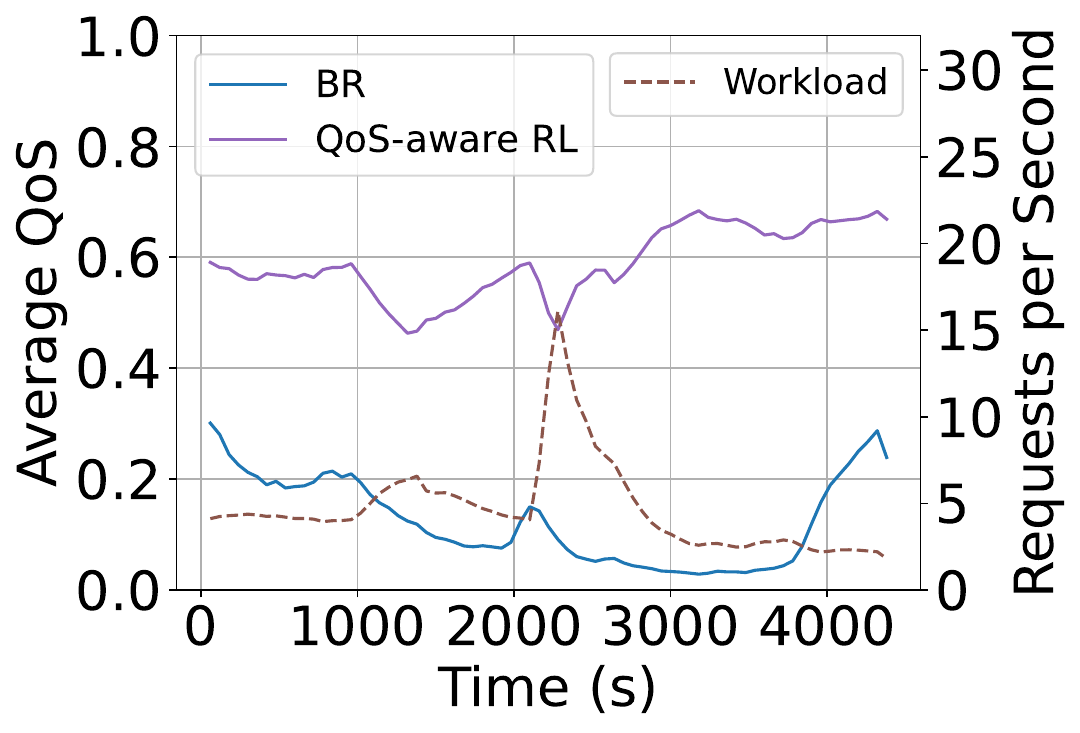}
    }
    \hfil
    \subfloat[RR vs QoS-aware RL]{
		\includegraphics[width=0.23\textwidth]{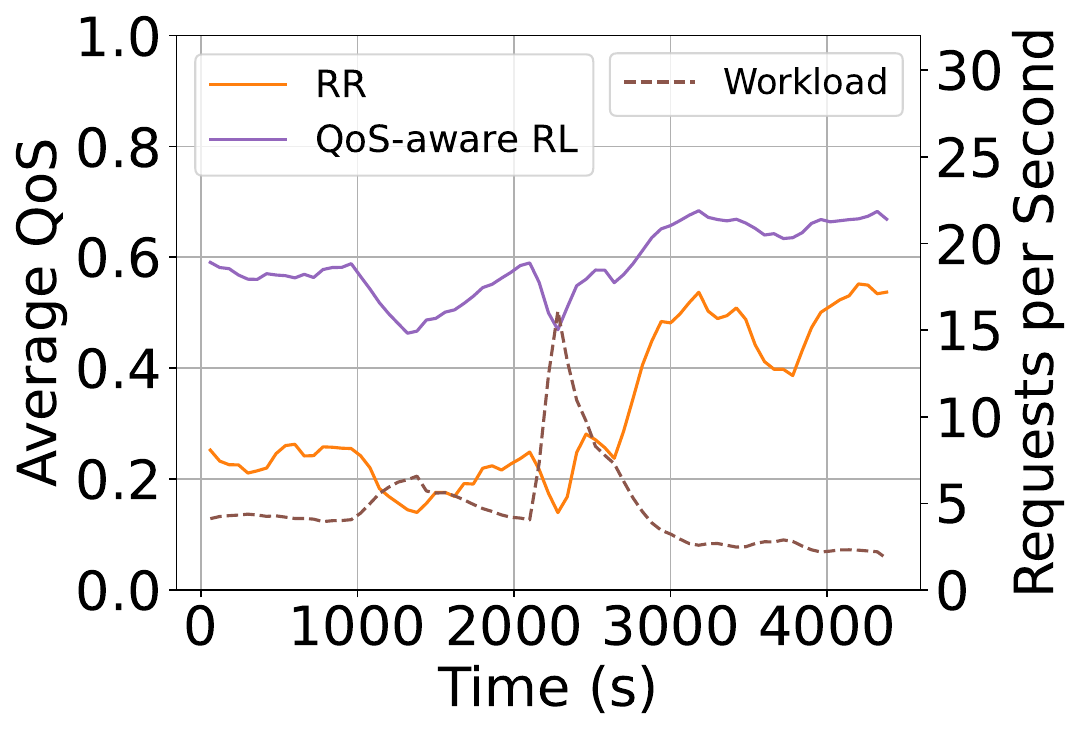} 
    }
    \hfil
    \subfloat[SQF vs QoS-aware RL]{
		\includegraphics[width=0.23\textwidth]{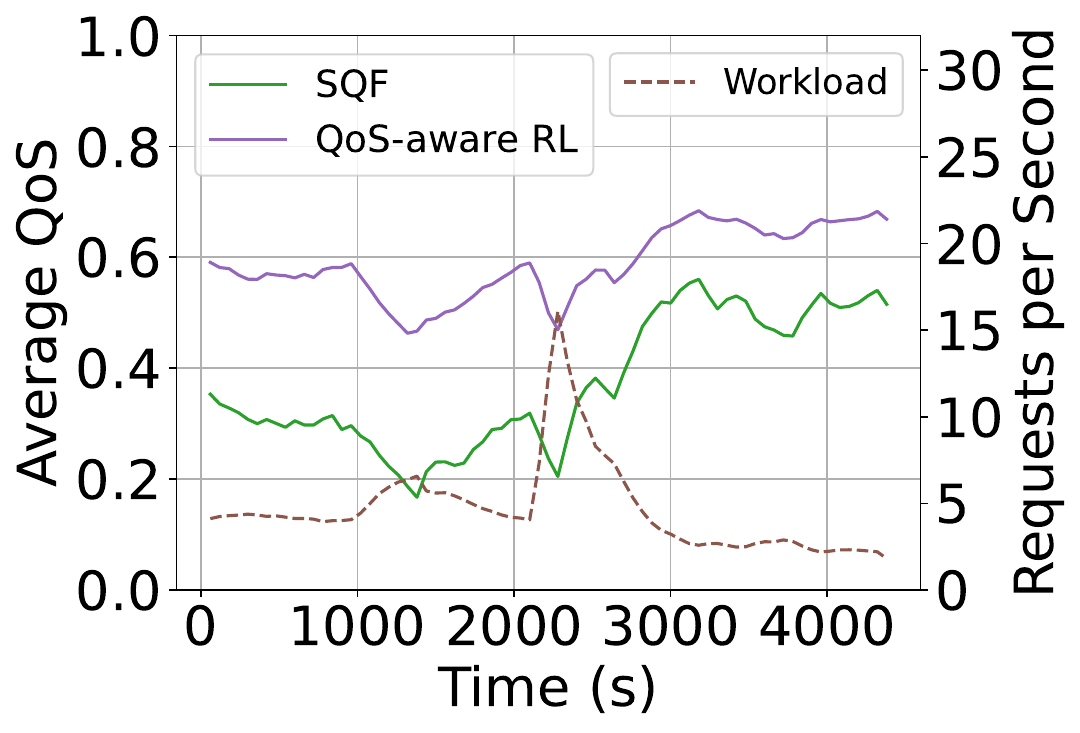} 
    }
    \hfil
    \subfloat[Baseline RL vs QoS-aware RL]{
		\includegraphics[width=0.23\textwidth]{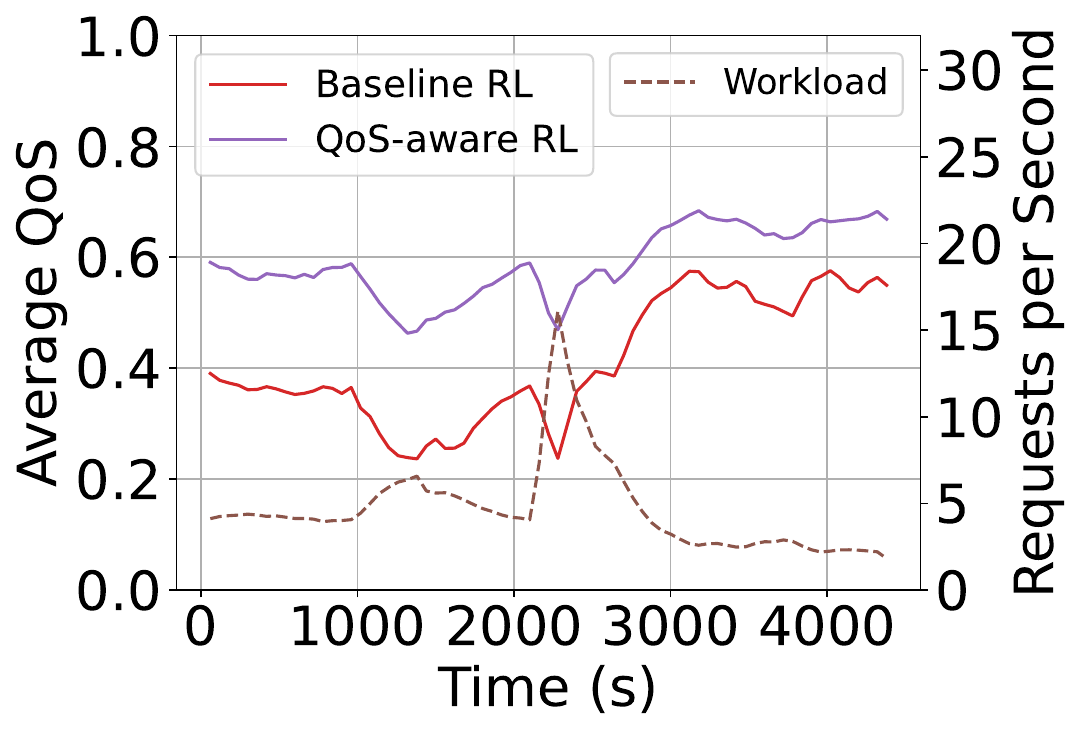} 
    }

    \caption{Average QoS for the long-running process with $N$=6 edge experts under real-world LLM workloads.}
    \label{exp:qos_long_term}
\end{figure*}

\begin{figure*}[t!]
    \centering
    \subfloat[BR vs QoS-aware RL]{
		\includegraphics[width=0.23\textwidth]{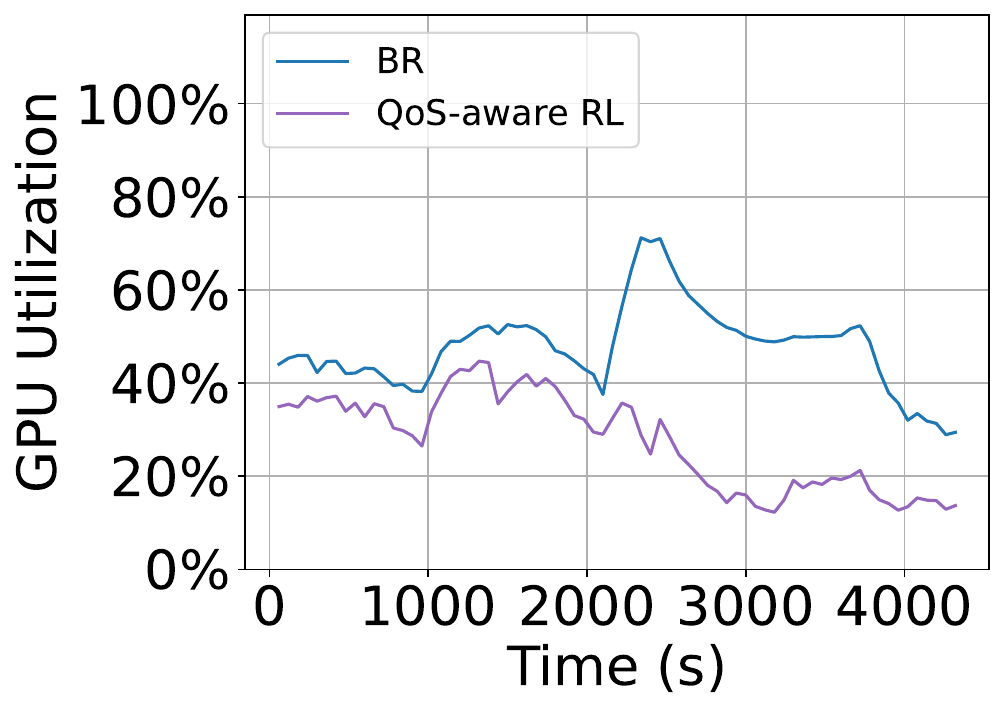}
    }
    \hfil
    \subfloat[RR vs QoS-aware RL]{
		\includegraphics[width=0.23\textwidth]{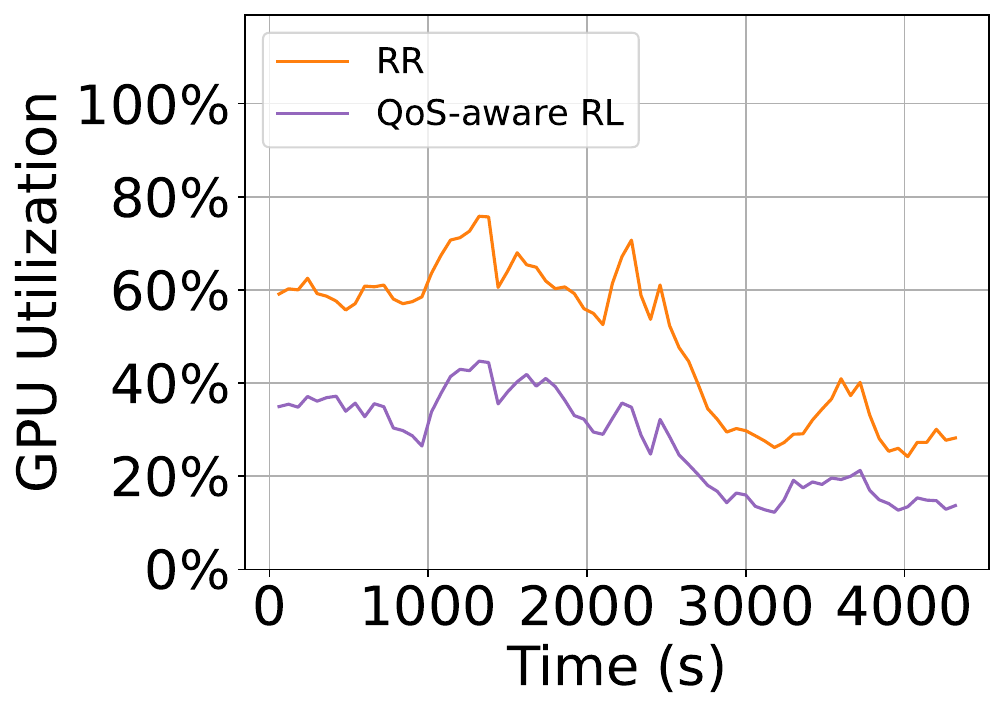} 
    }
    \hfil
    \subfloat[SQF vs QoS-aware RL]{
		\includegraphics[width=0.23\textwidth]{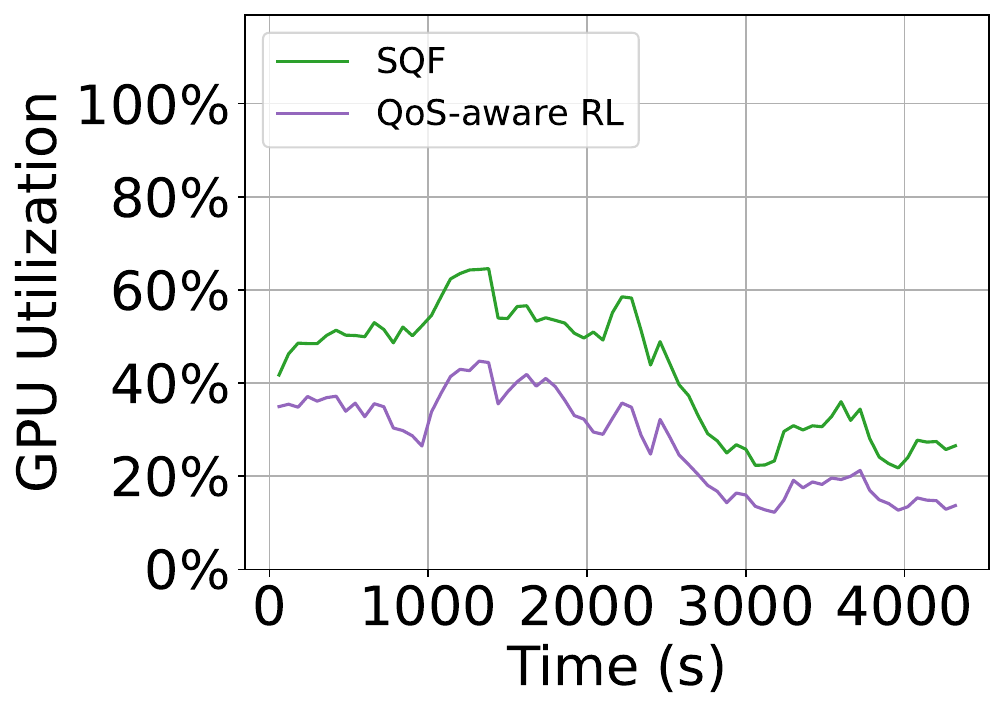} 
    }
    \hfil
    \subfloat[Baseline RL vs QoS-aware RL]{
		\includegraphics[width=0.23\textwidth]{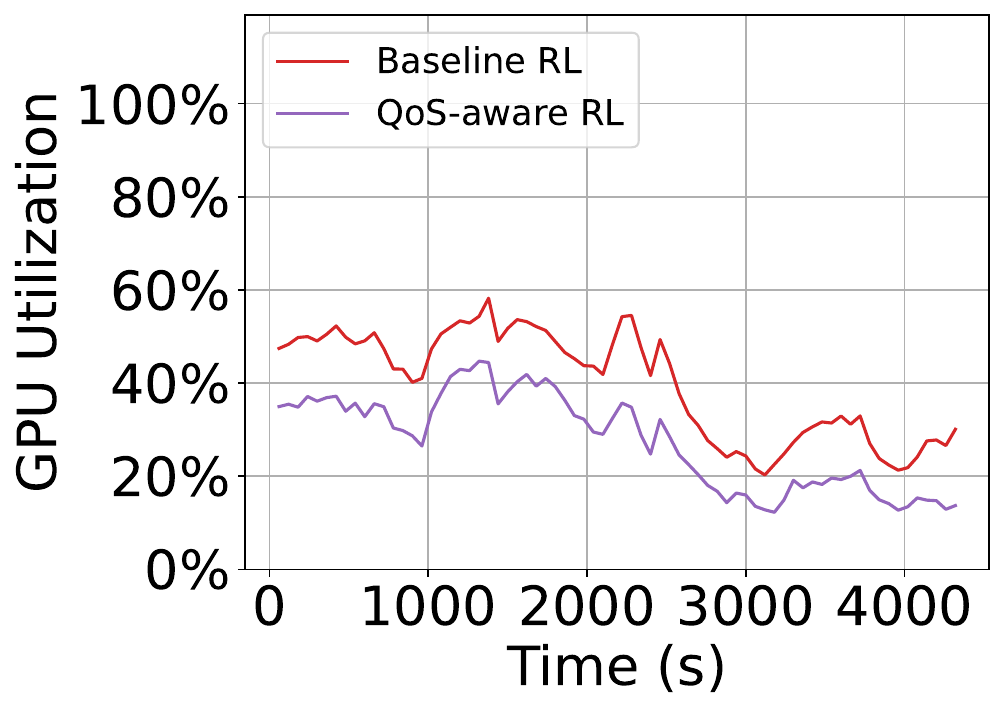} 
    }

    \caption{GPU usage for the long-running process with $N$=6 edge experts under real-world LLM workloads.}
    \label{exp:gpu_long_term}
\end{figure*}

\paragraph{End-to-end Latency Analysis} The end-to-end latency measurement encompasses the complete processing pipeline from user request arrival at the eAP to final response delivery, including four key components: (i) communication latency among the eAP and edge experts, (ii) routing latency, (iii) waiting delay at the assigned edge expert, and (iv) inference latency for response generation. As shown in Figure \ref{exp:e2e_latency}, we conduct a comprehensive analysis of the end-to-end latency with $N$=6 edge experts under Poisson workloads with $\lambda$=5. The communication latency remains below 1ms at a bandwidth of 1 Mbps, as only small-volume data (text-based user requests and small raw system state features) is involved, making it negligible in our analysis. Although our proposed algorithm introduces an additional 5ms of routing latency, as measured in experiments, this overhead accounts for only a small fraction of the total end-to-end latency, preserving its suitability for real-time routing applications. Additionally, the results reveal that LLM generation latency dominates the end-to-end latency profile. This observation motivates our focus on per-token latency measurement and optimization. Meanwhile, our proposed method outperforms baseline approaches in terms of latency per token, thereby achieving a substantial reduction in end-to-end latency and achieving improvements of at least 21.34\% over baselines.

\paragraph{Different Number of Edge Experts $N$} Given the constrained computational resources in edge environments and the high demand for LLM inference, the scale of edge experts is typically kept within a moderate range \cite{feng2024graphrouter,zhao2024eagle,hu2024routerbench}. To verify the scalability of our proposed algorithm, we scale the number of edge experts $N$ ranging from 3 to 12 while keeping other settings unchanged. As shown in Figure \ref{exp:varying_N}, our method consistently achieves the highest average QoS and the lowest average latency per token compared to baseline approaches.  Moreover, as $N$ increases and more edge experts are involved in service provision, the system benefits from greater computational resources. Consequently, our QoS-aware router delivers better performance in terms of both average QoS and average latency per token, showcasing its scalability concerning the number of edge experts.

\paragraph{Different Request Arrival Rates $\lambda$} To investigate the impact of workload intensity, we vary the request arrival rates $\lambda$ under Poisson workloads while keeping other settings unchanged. As shown in Figure \ref{exp:varying_lam}, the average QoS declines as workload intensity increases and computational resources become constrained. However, our proposed algorithm demonstrates a significantly slower degradation in QoS compared to the baselines, indicating its superior ability to make effective routing decisions even under high workload conditions. Moreover, our proposed approach maintains a more stable average latency per token across all workload intensities due to its explicit consideration of the latency requirements in decision-making. In conclusion, our proposed algorithm achieves at least a 34.11\% improvement in average QoS and a 4.17\% reduction in average latency per token compared to the baselines, highlighting its robustness and adaptability in dynamic and high workload conditions.

\paragraph{Different Latency Requirements $L$} To thoroughly investigate the influence of latency requirements, we conduct experiments by adjusting the system latency requirement $L$ while keeping all other experimental settings fixed. As shown in Figure \ref{exp:varying_L}, our proposed algorithm exhibits a slower decline in the average QoS compared to the baselines as the latency requirement becomes more stringent. Moreover, baseline methods fail to adapt effectively to varying latency requirements, with their average latency per token remaining largely unchanged across different values of $L$. In contrast, our proposed algorithm successfully adapts to stricter latency requirements and maintains lower average latencies per token that closely follow the target latency constraint. This superior adaptability can be attributed to the design of our reward function, which is tailored to accommodate diverse latency requirements, enabling our algorithm to perform robustly under varying latency constraints. In conclusion, our proposed algorithm achieves at least an 18.31\% improvement in the average QoS and a 4.63\% reduction in the average latency per token compared to the baselines across all tested latency requirements, demonstrating its adaptability and effectiveness in handling stringent latency constraints.

\paragraph{Long-running Process Visualization} To better understand the stability of our proposed algorithm under highly volatile workloads, we visualize the changes in the average QoS over time. Figure \ref{exp:qos_long_term} shows the average QoS of our proposed algorithm with $N$=6 edge experts under the real-world LLM workloads. We observe that our proposed algorithm stably outperforms baselines in such real-world long-running workloads. These results indicate that our trained router can be stably deployed online with multiple edge experts serving. Additionally, to evaluate the computational resource efficiency of our proposed method, we also look into the GPU usage of our proposed algorithm with $N$=6 edge experts under real-world LLM workloads. Figure \ref{exp:gpu_long_term} shows that our proposed algorithm strikes a good balance between the GPU memory efficiency and the overall QoS.

\subsection{Ablation Study}

\begin{figure}[t!]
    \centering
    \includegraphics[width=0.40\textwidth]{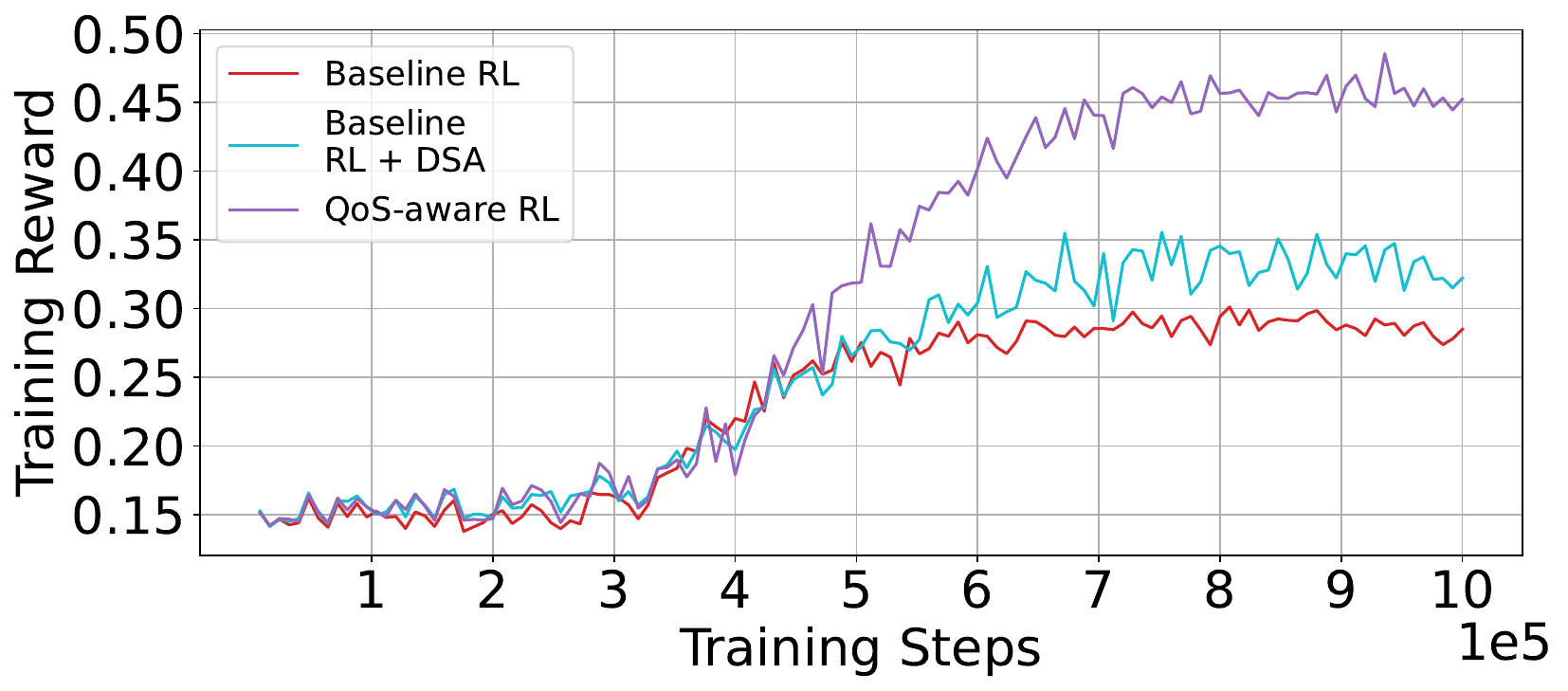}
	\label{fig:train_process}
    \caption{Training process with $N$=6 edge experts under Poisson workloads with $\lambda$=5.}
    \label{exp:train_process}
\end{figure}

\begin{figure}[t!]
    \centering
    \subfloat[Average QoS]{
		\includegraphics[width=0.23\textwidth]{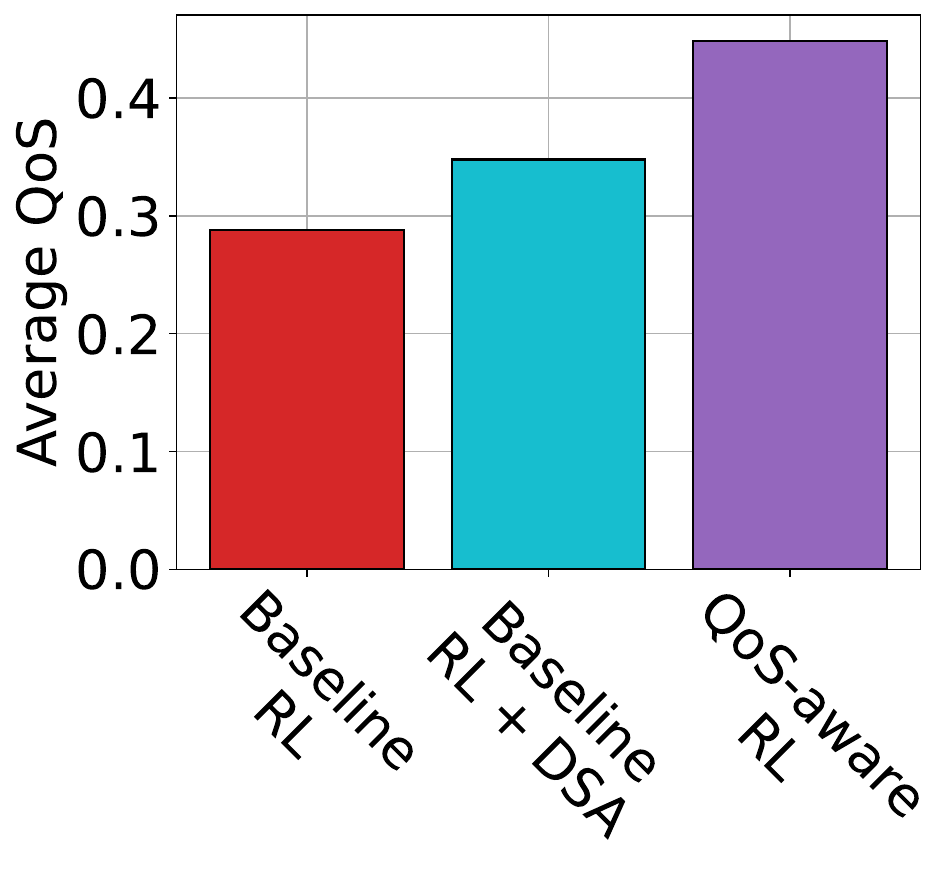}
		\label{fig:qos_ablation_study}
    }
    \hfil
    \subfloat[Average latency per token]{
		\includegraphics[width=0.23\textwidth]{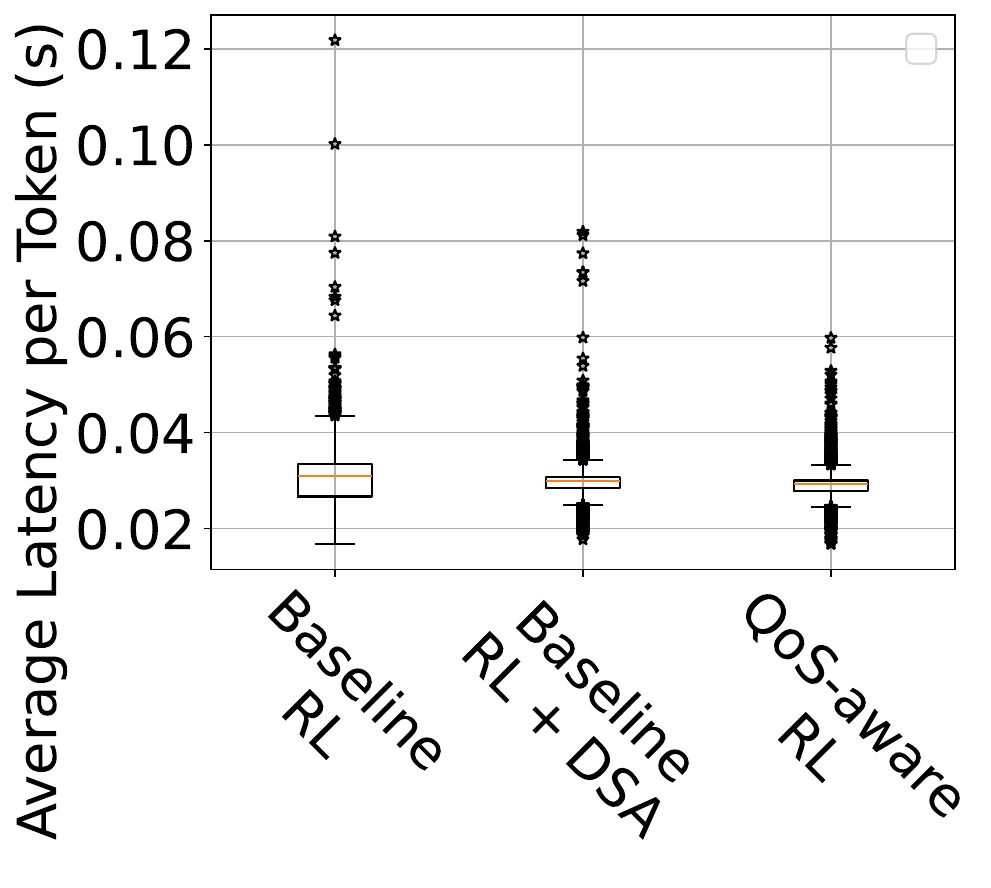} 
		\label{fig:latency_ablation_study}
    }

    \caption{Ablation study of dynamic state abstraction and QoS-aware reward with $N$=6 edge experts under Poisson workloads with $\lambda$=5.}
    \label{exp:ablation_study}
\end{figure}

\paragraph{Effectiveness of Dynamic State Abstraction and QoS-aware Reward} To evaluate the effectiveness of dynamic state abstraction (DSA) and QoS-aware reward, we conduct comparative experiments with three algorithm configurations as follows:
\begin{itemize}
    \item \textbf{Baseline RL.} This algorithm employs raw expert-level features without dynamic state abstraction and utilizes the baseline reward function described in Section VI-A.
    \item \textbf{Baseline RL + DSA.} This baseline algorithm integrates dynamic state abstraction while maintaining the same reward function as Baseline RL.
    \item \textbf{QoS-aware RL.} Our proposed algorithm combines both dynamic state abstraction and QoS-aware reward.
\end{itemize}
As shown in Figure \ref{exp:train_process}, we visualize the training process with $N$=6 edge experts under Poisson workloads with $\lambda$=5. Our QoS-aware router achieves superior convergence to a higher reward value within 1 million training steps. Both components demonstrate critical roles in enhancing learning efficiency and convergence capability, as evidenced by the significant performance gap compared to baseline approaches. To further quantify their contributions, we perform ablation studies as depicted in Figure \ref{exp:ablation_study}. The Baseline RL + DSA approach achieves a 17.27\% improvement in average QoS and a 3.22\% reduction in token latency compared to the Baseline RL approach. This significant performance gain demonstrates that our dynamic state abstraction effectively captures the system dynamics by providing compact and fine-grained request representations that reveal critical computational resource utilization patterns. Additionally, our proposed QoS-aware RL approach further enhances performance with an additional 22.37\% QoS improvement while maintaining 2.15\% latency reduction. This improvement stems from our action impact estimator, which precisely quantifies how each routing decision affects overall QoS, thereby enabling the designed QoS-aware reward function to guide the DRL agent effectively and mitigate potential latency violations. In conclusion, these results underscore the synergistic benefits of combining both components for optimal overall QoS.

\begin{figure}[t!]
    \centering
    \subfloat[Average QoS]{
		\includegraphics[width=0.23\textwidth]{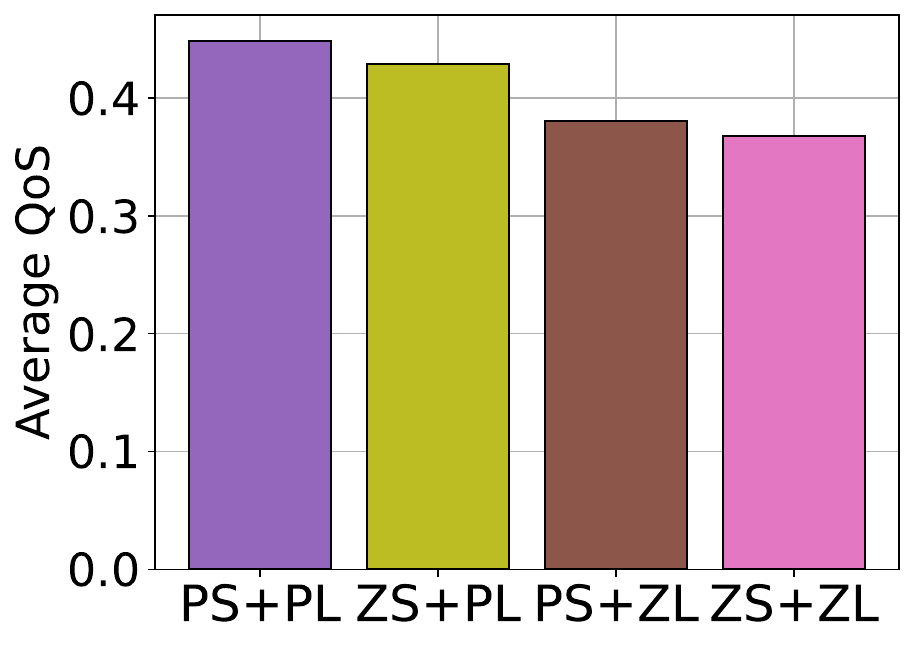}
		\label{fig:qos_poisson_5.0_6_predictor_ablation}
    }
    \hfil
    \subfloat[Average latency per token]{
		\includegraphics[width=0.23\textwidth]{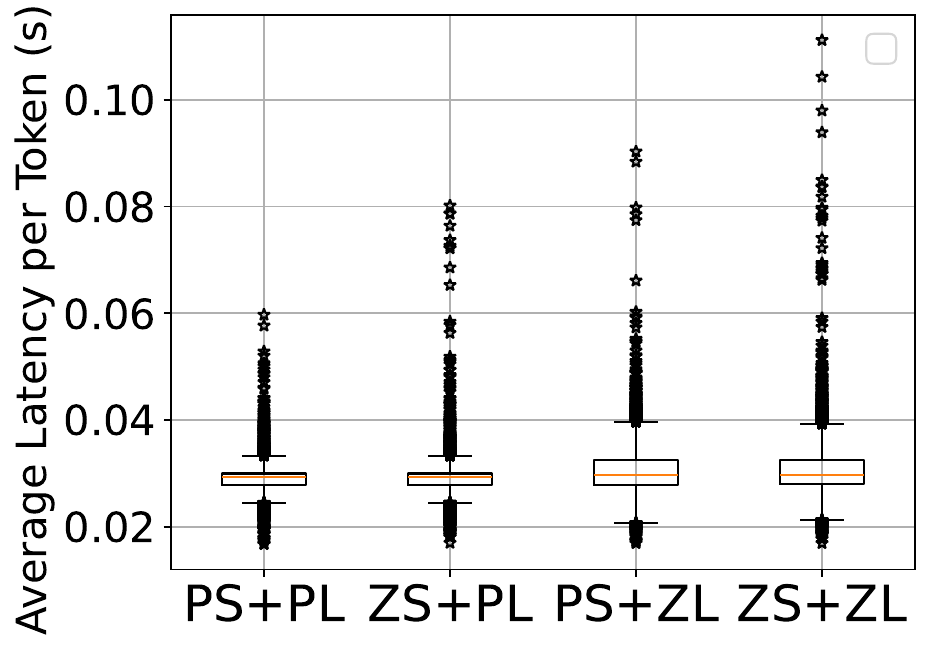} 
		\label{fig:latency_poisson_5.0_6_predictor_ablation}
    }
    \caption{Ablation study of generation score and output length predictors with $N$=6 edge experts under Poisson workloads with $\lambda$=5.}
    \label{exp:predictor_ablation}
\end{figure}

\paragraph{Effectiveness of Generation Score and Output Length Predictors} To evaluate the effectiveness of generation scores and output length predictors, we design a series of experiments to compare the performance of different combinations. Specifically, we compare the following combinations,
\begin{itemize}
    \item \textbf{PS+PL.} Our proposed algorithm that use \textbf{P}redicted generation \textbf{S}core and \textbf{P}redicted output \textbf{L}ength in the raw feature.
    \item \textbf{ZS+PL.} It replace the raw feature with \textbf{Z}ero generation \textbf{Z}core and \textbf{P}redicted output \textbf{L}ength.
    \item \textbf{PS+ZL.} It replace the raw feature with \textbf{P}redicted generation \textbf{Z}core and \textbf{Z}ero output \textbf{L}ength.
    \item \textbf{ZS+ZL.} It replace the raw feature with \textbf{Z}ero generation \textbf{Z}core and \textbf{Z}ero output \textbf{L}ength.
\end{itemize}
As shown in Figure \ref{exp:predictor_ablation}, our results show that employing our predictors still leads to a 17.94\% improvement in the average QoS and 1.33\% reduction in the average latency per token over no predictive information is available. Note that even with no predictive information, our proposed algorithm still outperforms baseline methods by at least 21.74\%, demonstrating both the importance of the predictors and the inherent resilience of the proposed algorithm to imperfect predictions. These findings highlight that while the predictors significantly enhance performance, the DRL agent’s adaptive routing is guided by both real-time and predicted system states, ensuring resilience against prediction uncertainties.

\section{Conclusion}
\label{sec:conclusion}
To maximize the long-term QoS for user requests, we propose a novel DRL-based QoS-aware LLM routing algorithm designed to achieve optimized routing under dynamic workloads. Due to the dynamic nature of the global state, we propose a dynamic state abstraction technique with a HAN to efficiently abstract the dynamic global state features. Besides, we propose an action impact estimator and a tailored reward function to guide the DRL agent in maximizing overall QoS and preventing latency requirement violations. Experiments demonstrate that our proposed algorithm can improve average QoS by up to 35.78\% compared to baselines under both Poisson and real-world LLM workloads.

\bibliographystyle{IEEEtran}
\bibliography{IEEEabrv, references}

\end{document}